\documentclass[twocolumn,amsmath,amssymb,aps,prl,floatfix,superscriptaddress]{revtex4}
\usepackage{graphicx}
\usepackage{braket}
\usepackage{amsmath}
\usepackage{enumerate}
\usepackage{color}
\DeclareMathOperator{\Tr}{Tr}

\begin{document}
\title{Machine Learning Assisted Many-Body Entanglement Measurement}

\author{Johnnie Gray}
    \affiliation{Department of Physics and Astronomy, University College London, Gower Street, London WC1E 6BT, United Kingdom}
    \email{john.gray.14@ucl.ac.uk}
\author{Leonardo Banchi}
    \affiliation{Department of Physics and Astronomy, University College London, Gower Street, London WC1E 6BT, United Kingdom}
    \email{benkj.it@gmail.com}
\author{Abolfazl Bayat}
    \affiliation{Institute of Fundamental and Frontier Sciences, University of Electronic Science and Technology of China, Chengdu 610051, China  }
    \affiliation{Department of Physics and Astronomy, University College London, Gower Street, London WC1E 6BT, United Kingdom}
    \email{abolfazl.bayat@ucl.ac.uk}
\author{Sougato Bose}
    \affiliation{Department of Physics and Astronomy, University College London, Gower Street, London WC1E 6BT, United Kingdom}
    \email{s.bose@ucl.ac.uk}
\date{\today}

\begin{abstract}
Entanglement not only plays a crucial role in quantum technologies, but is key to our understanding of quantum correlations in many-body systems.
However, in an experiment, the only way of measuring entanglement in a generic mixed state is through reconstructive quantum tomography, requiring an exponential number of measurements in the system size.
Here, we propose a machine learning assisted scheme to measure the entanglement between arbitrary subsystems of size $N_A$ and $N_B$, with $\mathcal{O}(N_A + N_B)$ measurements, and without any prior knowledge of the state.
The method exploits a neural network to learn the unknown, non-linear function relating certain measurable moments and the logarithmic negativity.
Our procedure will allow entanglement measurements in a wide variety of systems, including strongly interacting many body systems in both equilibrium and non-equilibrium regimes.
\end{abstract}

\maketitle
\emph{Introduction. -- }
Entanglement is a key property for many emerging quantum technologies~\cite{rao1945information,shor1999polynomial,harrow2009quantum,bennett1993teleporting,bennett1992communication,ekert1991quantum}, but also is essential for understanding the structure of strongly correlated many-body systems~\cite{amico2008entanglement,schollwock2011density}.
Despite its paramount importance, only for the very limited case of a bipartition of a pure state can the entanglement, quantified by subsystem entropy, be measured in an efficient and state-independent way~\cite{horodecki2002method}.
There are multiple proposals to carry out such a scheme in various physical systems, such as optical lattices~\cite{daley2012measuring,alves2004multipartite}, quantum dot arrays~\cite{banchi2016entanglement} and Gaussian systems~\cite{weedbrook2012gaussian}.
Recently, some of these have also been experimentally realised in simulated spin chains, for example in cold atoms \cite{islam2015measuring} and photonic chips~\cite{pitsios2016photonic}.
Nonetheless, pure states are very rare: they are not only difficult to prepare in realistic situations, but also difficult to maintain in the presence of an environment.
For example, just consider the entanglement between: (i) optical modes traversing fibres, crucial for quantum communication; (ii) spatially separated parts of an extended many-body pure state, important for characterizing long range entanglement~\cite{reznik2003entanglement,bayat2010negativity,alkurtass2016entanglement,gray2017many,banchi2009finite,roscilde2004studying}; (iii) two systems in a thermal state -- in none of the above cases, ironically, can the entanglement entropy quantify the entanglement.
Witnesses do exist for specific forms of entanglement, but these are state-dependent and provide only a simple yes/no answer~\cite{HORODECKI19961,terhal2000bell,lewenstein2000optimization}, or bounds on the quantity of entanglement~\cite{brandao2005quantifying,audenaert2006correlations,cavalcanti2006estimating,eisert2007quantitative,guhne2007estimating,guhne2008lower,carteret2016estimating}.
However, the crucial task of being able to accurately measure entanglement for mixed states in an experimental setting remains open.

While for pure states bipartite entanglement is uniquely defined by the entropy of the subsystems, for mixed states the landscape is far more complex~\cite{horodecki2009quantum,plenio2007introduction}.
Aside from isolated special cases such as two qubit states~\cite{wootters1998entanglement} and bosonic Gaussian states~\cite{weedbrook2012gaussian,wolf2004gaussian}, only the (logarithmic) negativity~\cite{zyczkowski1998volume,lee2000partial,vidal2002computable,plenio2005logarithmic} is a computationally tractable quantity~\cite{huang2014computing}.
It bounds the distillable entanglement and teleportation capacity \cite{vidal2002computable}, and is a pivotally important quantity to estimate for both quantum technologies~\cite{plenio2007introduction,horodecki2009quantum,lanyon2016efficient} and condensed matter systems~\cite{calabrese2012entanglement}.
Nonetheless, there is no state-independent observable that can measure the logarithmic negativity, and thus its experimental measurement requires full state tomography~\cite{d2003quantum} --- demanding, in general, an exponential number of measurements in the system size.
Recently, polynomial tomography schemes have emerged, such as tensor networks for lowly entangled states~\cite{cramer2010efficient,lanyon2016efficient}, or  breakthroughs in neural network state reconstruction~\cite{torlai2018neural,torlai2018latent}.
However, these may be insufficient for estimating entanglement, since many entanglement measures, such as the logarithmic negativity, are not continuous~\cite{bengtsson2006geometry}.
Namely, even if reconstructed state $\rho_r$ approximates actual state $\rho$ closely, the two may have significantly different negativities~\cite{asymptoticContinuityNote}.

Here, we put forward a machine learning assisted scheme for accurately estimating the logarithmic negativity in a completely general and realistic setting, using an efficient number of measurements -- scaling polynomially with system size.
Our estimator works for a wide range of states, and is
remarkably accurate for highly entangled states.
Our method is based on measuring a finite number of moments of a partially transposed density matrix~\cite{carteret2005noiseless,cai2008novel,Bartkiewicz2014method} from which we extract the entanglement negativity using machine learning.
This {\it direct} estimation of negativity avoids approximate state reconstruction~\cite{cramer2010efficient,lanyon2016efficient,torlai2018neural,torlai2018latent},
and represents a new front in applying classical machine learning to quantum problems~\cite{carrasquilla2017machine,carleo2017solving,schollwock2011density,hentschel2010machine,banchi2016quantum}.
Moreover, we propose a new method for measuring those moments, beyond~\cite{carteret2005noiseless,cai2008novel,Bartkiewicz2014method}, which is experimentally feasible in the many-body setting, since the individual building blocks have already been demonstrated in solid state~\cite{petta2005coherent} and cold atoms~\cite{islam2015measuring}.

\begin{figure*}[t]
    \centering
    \includegraphics[width=0.8\linewidth]{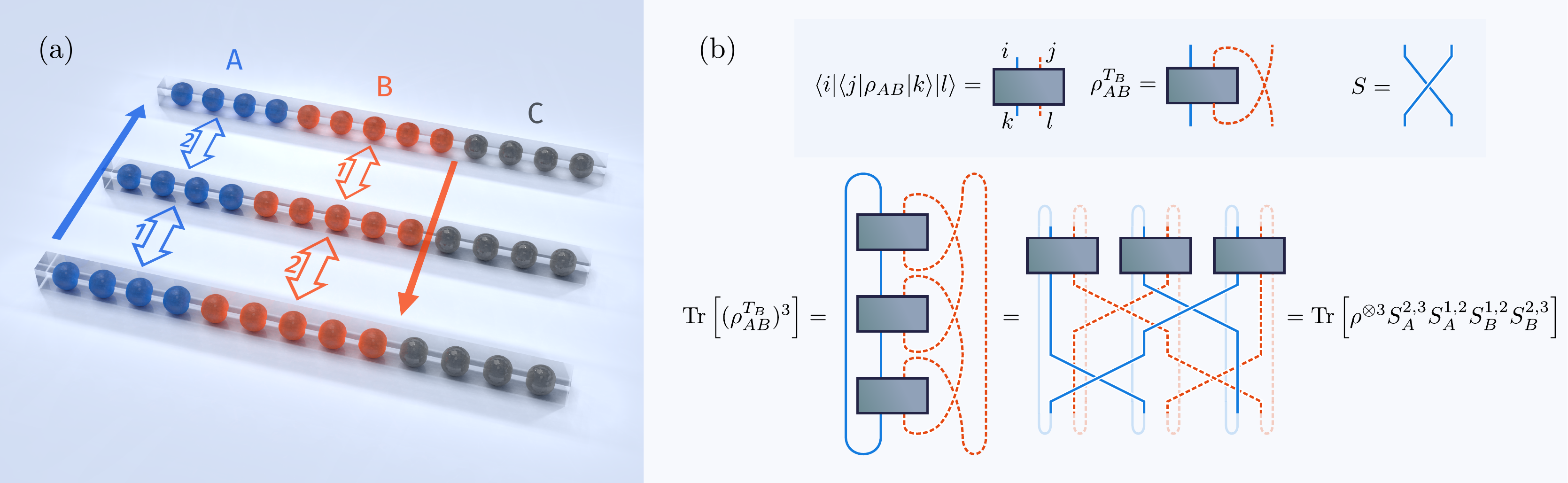}
    \caption{
    \textbf{Schematics:}
    (a) Example measurement set-up for the moments, $\mu_m = \Tr\left[(\rho_{AB}^{T_B})^m\right]$, here for $m=3$, from which one can extract the logarithmic negativity $\mathcal{E}$ between $A$ and $B$.
    The generic mixedness of $\rho_{AB}$ could arise from entanglement with environment $C$.
    Here the subsystems contain $N_A$, $N_B$ and $N_C$ particles respectively.
    The scheme involves three copies of the original system, and two counter propagating sets of measurements on $A$ and $B$, ordered by the shown numbers, with direction depicted by the filled arrows.
    (b) Diagrammatic proof (for $m=3$) of the equivalence between the moments $\mu_m$ and expectation of two opposite permutations (decomposed as swaps) on $A$ and $B$ -- from which a measurement scheme can be derived.
    }
    \label{fig:schematics}
\end{figure*}


\emph{Logarithmic Negativity.} ---
Logarithmic negativity~\cite{zyczkowski1998volume,lee2000partial,vidal2002computable,plenio2005logarithmic} for a generic mixed state $\rho_{AB}$ quantifies the entanglement between subsystems $A$ and $B$.
It is defined as:
\begin{equation} \label{eq:negativity}
    \mathcal{E}
    = \log_2 \left| \rho_{AB}^{T_A}\right|
    = \log_2 \left| \rho_{AB}^{T_B}\right|
    = \log_2 \sum_k |\lambda_k |
\end{equation}
with $| \cdot |$ the trace norm, $\rho_{AB}^{T_X}$ the partial transpose with respect to subsystem $X$, and $\{\lambda_k\}$ the eigenvalues of $\rho_{AB}^{T_X}$.
Because of the non-trivial dependence of $\mathcal{E}$ on $\rho_{AB}$, there is no state-independent observable that can measure it --- generally demanding full state tomography.
The $\{\lambda_k\}$ are the roots of the characteristic polynomial, $P(\lambda)  {=} \det(\lambda - \rho_{AB}^{T_B}) {=} \sum_{n}c_n\lambda^{n}$, where each $c_n$ is a polynomial function of the partially transposed moments:
\begin{equation} \label{eq:moments}
    \mu_m = \Tr\left[(\rho_{AB}^{T_B})^m\right]
    =
    \sum_k \lambda_k^m.
\end{equation}
In this way, full information about the spectrum $\{\lambda_k\}$ is contained in $\{\mu_m\}$.
It is known that these measuring these moments is technically possible using $m$ copies of the state and controlled swap operations~\cite{carteret2005noiseless}.
However, even if these experimentally challenging operations were available, the problem of extracting $\{\lambda_k\}$ from the moments is notoriously ill-conditioned~\cite{viano1991solution}, with a closely related problem being described as numerically catastrophic.
Alongside this, an exponential number of moments respective to the size of $AB$ are needed to exactly solve the equations.
On the other hand, to estimate the logarithmic negativity, a precise knowledge of all $\lambda_k$ is not required.
Since $-\frac{1}{2} \le \lambda_k \le 1$ for all $k$ \cite{rana2013negative} and $\sum_k \lambda_k = 1$, generically, the magnitude of the moments quickly decreases with $m$, with the first few carrying the most information.
Backing up this intuition, we will show that the moments required, $\{\mu_m : m \le M\}$, to accurately estimate the entanglement can number as few as $M=3$.
We do this by employing machine learning to \emph{directly} map moments to logarithmic negativity, avoiding reconstruction of the spectrum or state.
Note that $\mu_0$ is simply the dimension of the systems Hilbert space, while $\mu_1=1$ in all cases.
Additionally, it can be easily shown that $\mu_2$ is equal to the purity of the state $=\Tr\left[\rho_{AB}^2\right]$, and as such, $M\ge 3$ is needed to extract any information about $\mathcal{E}$.
In this sense our method is optimal in terms of number of copies.

\emph{Measuring the Moments of $\rho_{AB}^{T_B}$.} --
The method for measuring the moments proposed in~\cite{carteret2005noiseless} based on 3-body controlled swaps is practically challenging in a many-body set-up where natural interactions are two-body.
A simpler protocol, for 4 moments only, was provided in~\cite{Bartkiewicz2014method}.
Here, we show that any moment in Eq.~\eqref{eq:moments} can be measured using only SWAP-operators between the \emph{individual} constituents of the
$m$ copies of the state $\rho_{AB}$, namely $\rho_{AB}^{\otimes m} = \bigotimes_{c=1}^m \rho_{A_c B_c}$.
This general set-up is shown in Fig.~\ref{fig:schematics}(a), where the mixedness of $\rho_{AB}$ arises from possible entanglement with a third system $C$, such that $\rho_{AB} = \Tr_C\ket{\Psi_{ABC}}\bra{\Psi_{ABC}}$ with $\ket{\Psi_{ABC}}$ being a pure tripartite state.  The first step is to write the matrix power as an expectation of a permutation operator, similar to Ref.~\cite{ekert2002direct,horodecki2002method}, but here on the partially transposed copies:
\begin{align}\label{eq:moments-and-perms}
    \mu_m
    &= \Tr\left[
    \left (
    \bigotimes_{c=1}^m
    \rho_{A_c B_c}^{T_{B_c}}
    \right)
    \mathbb{P}^{m}
    \right]
    \cr &=
    \Tr\left[
    \left (
    \bigotimes_{c=1}^m
    \rho_{A_c B_c}
    \right)
    (\mathbb{P}^{m})^{T_B}
    \right]~,
\end{align}
where $\mathbb{P}^{m}$ is any linear combination of cyclic permutation operators of order $m$ and the second line makes use of the identity $\Tr(\rho_{AB}^{T_B}O) {=} \Tr(\rho_{AB} O^{T_B})$, valid for any operator $O$.
A schematic of the equality in Eq.~\eqref{eq:moments-and-perms} for $m=3$ is shown in Fig.~\ref{fig:schematics}(b).
In the appendix~\footnote{The appendix includes references
~\cite{banchi2016entanglement,blaizot1986quantum,puddy_multiplexed_2015,veldhorst2016silicon,petta2005coherent,islam2015measuring,daley2012measuring,mead1984maximum,viano1991solution,han2016approximating,eisler2015partial,coser2016towards,aubrun2012partial,fukuda2013partial,schollwock2011density,bergstra2013making,chollet2015keras,lanyon2016efficient}} we provide a choice of $\mathbb{P}^{m}$ with a neat operational meaning, both for spin and bosonic systems.
For spin lattices,  our  choice of $\mathbb{P}^m$ to measure the moments $\mu_m$ results to the following steps in practice:
(i) prepare $m$ copies of the state $\rho_{AB}$;
(ii) sequentially measure a `forward' sequence of adjacent swaps, $S_A^{c, c+1}$ between neighbouring copies of system $A$ from $c=1$ to $m-1$;
(iii) sequentially measure a `backward' sequence of adjacent swaps, $S_B^{c, c-1}$ between neighbouring copies of system $B$ from $c=m$ to $2$;
(iv) repeat these steps in order to yield an expectation value.
This procedure is also depicted for $m = 3$ in Fig.~\ref{fig:schematics}(a).
For bosonic lattices, our procedure corresponds to the following steps:
(i) prepare $m$ copies of the state $\rho_{AB}$;
(ii) Perform  `forward' Fourier transforms between modes in different copies for each site in $A$ --
this can be achieved using a series of beam splitters \cite{reck1994experimental};
(iii) Perform `backwards' (reverse) Fourier transform between modes in different copies
for each site in $B$, via reverse beam splitter transformations;
(iv) Measure the boson occupation numbers $n_{j,c}$ on all sites $j \in\{A,B\}$  and all copies $c$ to compute $\phi=e^{i\sum_{j\in\{A,B\},c} 2\pi c n_{j,c}/m}$.
(v) Repeat these steps to obtain the expectation value $\mu_m$ as an average of $\phi$.
Both procedures require $\mathcal{O}(N_A + N_B)$ measurements for each $m$ between $2$ and $M$, and are explained in detail in the appendix.
This is in stark contrast to tomography, which generically for qubit systems requires $2^{2(N_A + N_B)}$ measurement settings.

It is worth emphasizing the difference between our procedure, and recently proposed operational methods for measuring Renyi entropies~\cite{daley2012measuring,banchi2016entanglement,abanin2012measuring}.
First of all, Renyi entropies only quantify entanglement for pure states, and cannot be used in the more general mixed state scenario.
Secondly, while for entropies the operations are only performed on a single subsystem, here, one performs both `forward' and `backward' operations on two subsystems at once, as explained above.
Remarkably, even though partially transposed density matrices are generically un-physical, measurement of their moments is possible.


\bigskip
\emph{Machine Learning Entanglement.} --
We focus now on estimating the logarithmic negativity from the information contained in the moments, $\mu_m$.
One approach using only the even moments has been proposed in the quantum field theory literature~\cite{calabrese2012entanglement,de2015entanglement} by exploiting numerical extrapolation.
However, this method neglects the odd moments and generally requires a large number of moments and thus copies.
We have developed an alternative analytical method based on Chebyshev functional approximation, detailed in the appendix, which takes into account these odd moments.
Indeed with the same number of copies we find it produces more accurate estimates, and thus serves as a reference quantity.
The Chebyshev expansion is analytically tractable, and becomes accurate for large enough $M$, as is shown in the appendix.
Nonetheless, this expansion is based on a linear mapping between the moments and the negativity, despite this relationship being inherently non-linear.
Therefore it is natural to think that a non-linear transformation could be more optimal, and thus more efficient for smaller $M$ -- namely fewer copies.

Machine learning has recently emerged as a key tool for modelling an unknown non-linear relationship between sets of data.
In the supervised learning paradigm, one trains a model with a set of known inputs and their corresponding outputs.
Once trained, the model can then be used to predict the unknown output of new input data.
Here, we take the moments $\mu_m$ as the input and the logarithmic negativity $\mathcal{E}$ as the output.
Training is performed by taking a large set of states for which $\mu_m$ and $\mathcal{E}$ can be computed on a classical computer.
This model can then be used to predict $\mathcal{E}$ from a set of experimentally measured moments.
The experimental system under study motivates the choice of which training states to use, so that they share, for example, similar entanglement features.
Among the most successful machine learning algorithms for non-linear regression are supervised vector machines~\cite{cristianini2000introduction}, random decision forests~\cite{ho1998random}, and deep neural networks~\cite{rojas2013neural,schmidhuber2015deep}.
However, we have found that using the same training set for each, neural networks are superior when it comes to predicting logarithmic negativity for a wide range of states beyond the training set.
As we show with our numerical results, neural networks provide a very accurate method for extracting the logarithmic negativity with as few as $M=3$ copies.
The details of our neural network construction can be found in the appendix.

\begin{figure}[tb]
    \centering
    \includegraphics[width=\linewidth]{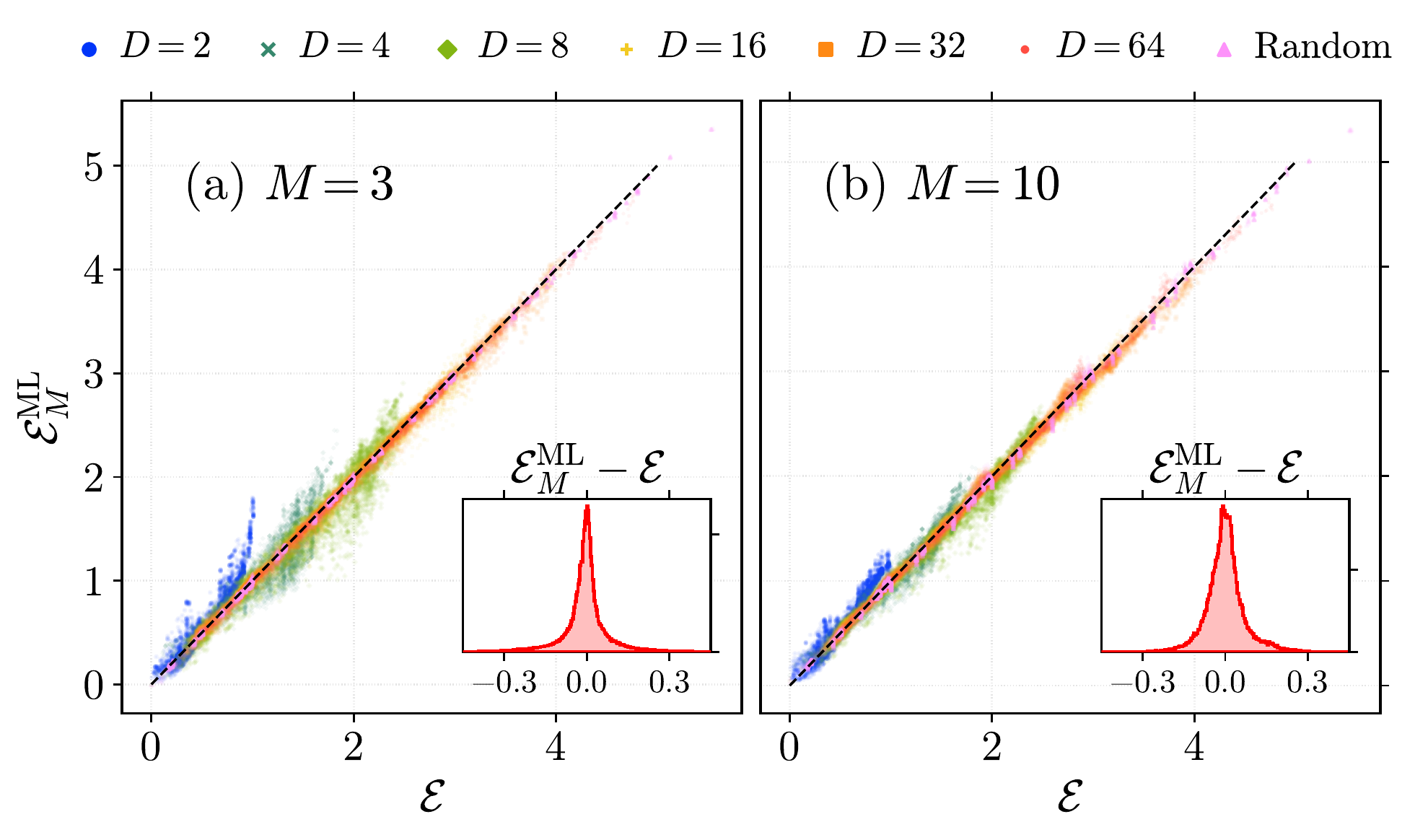}
    \caption{
    \textbf{Machine learning entanglement.}
    Estimated logarithmic negativity $\mathcal{E}^{\mathrm{ML}}_M$, using a machine learning vs. actual logarithmic negativity $\mathcal{E}$, for the same set of random states described in the main text.
    Training and prediction is performed using the moments $\mu_m$ generated from: (a) $M=3$ copies; (b) $M=10$ copies.
    The respective insets show the distribution of error, $\mathcal{E}^{\mathrm{ML}}_M - \mathcal{E}$.
    }
    \label{fig:ml-logneg-approx}
\end{figure}

\bigskip

\emph{Training with Random States.} --
In order to train a neural network, a set of suitable training states are required for which both the moments and logarithmic negativities are known.
From an entanglement perspective, relevant states in condensed matter physics can be classified as either area-law, or volume-law.
In the first case, the entanglement of a subsystem $A$ with the rest is proportional to the number of qubits along their boundary.
In the second, this entanglement is instead proportional to $N_A$, the number of qubits in $A$.
Area-law states arise as low energy eigenstates of local gapped Hamiltonians, with logarithmic corrections in critical systems.
Volume-law states however, are associated with the eigenstates found in the mid-spectrum, and as such arise in non-equilibrium dynamics, e.g. quantum quenches~\cite{barmettler2009relaxation,nanduri2014entanglement}.

Rather than concentrate on a training with a specific model system, we initially consider the very general case of random states.
To encompass both area- and volume-law states, we consider two classes of states $\ket{\Psi_{ABS}}$: (i) random generic pure states (R-GPS), e.g. sampled from the Haar measure, which typically have volume-law entanglement~\cite{popescu2006entanglement,hamma2012quantum}; (ii) random matrix product states (R-MPS) with fixed bond dimension, which satisfy an area-law by construction \cite{schollwock2011density}.
In order to generate a training set with a wide range of entanglement features, subsystem sizes, and mixedness, we perform the following procedure:
(i) For a fixed number of qubits $N$, take either a R-GPS, or R-MPS with bond dimension $D$.
(ii) Take different tri-partitions such that $N = N_A + N_B + N_C$, and for each calculate ${\mu_m}$ and $\mathcal{E}$ for $\rho_{AB}$.
(iii) Repeat for different random instances, while separately varying $N$ and $D$.
Further generation and training are provided in the appendix.

\bigskip

\emph{ Numerical Results for Random States.} --
To check the performance of our neural network estimator, we take the set of random states described in the previous section and split this data in two, one half for training the neural network model, and the other as `unseen' test data.
In Fig.~\ref{fig:ml-logneg-approx}(a) we plot the machine learning model's predictions, $\mathcal{E}^{\mathrm{ML}}_{M}$, for the test data, using only $M=3$ copies, in which a high degree of accuracy is achieved.
In the inset of Fig.~\ref{fig:ml-logneg-approx}(a), we plot a histogram of the errors $\mathcal{E}^{\mathrm{ML}}_M - \mathcal{E}$, which displays a very sharp peak at zero error with standard deviation $\sim 0.09$
A further improvement, particularly in outliers, is achieved by increasing the number of copies $M$ to 10, see Fig.~\ref{fig:ml-logneg-approx}(b), where the error standard deviation decreases to $\sim 0.07$
Regardless, the machine learning method is already very accurate for extracting entanglement using only three copies.
The machine learning approach works particularly well for large bond dimension and volume-law like states -- an important fact given that these are the exact cases where efficient tomography fails.
A more detailed discussion about sensitivity and ascribing errors to machine learning predictions can be found in the appendix.

\bigskip

\emph{ Numerical Results for Physical States. } --
We now consider the more realistic setting of quench dynamics in a many-body system.
We take a system of $N$ spin-1/2 particles with nearest neighbour Heisenberg Hamiltonian
$ H= J \sum_{i=1}^{N-1} \boldsymbol{\sigma}_i \cdot \boldsymbol{\sigma}_{i+1}$ with $J$ the interaction strength and
$\boldsymbol{\sigma}_{i} = (\sigma^x_i, \sigma^y_i, \sigma^z_i)$
the vector of Pauli matrices acting on site $i$.
The system is initialised in the (separable)
Neel-state $\ket{\Psi(0)} = \ket{\uparrow \downarrow \uparrow \dots}$.
As the chain unitarily evolves in time as $\ket{\Psi(t)} = e^{-i H t} \ket{\Psi(0)}$, it becomes entangled, with an effective MPS description whose bond dimension increases until the state is essentially volume-law~\cite{barmettler2009relaxation,nanduri2014entanglement}.

\begin{figure}[tb]
    \centering
    \includegraphics[width=0.9\linewidth]{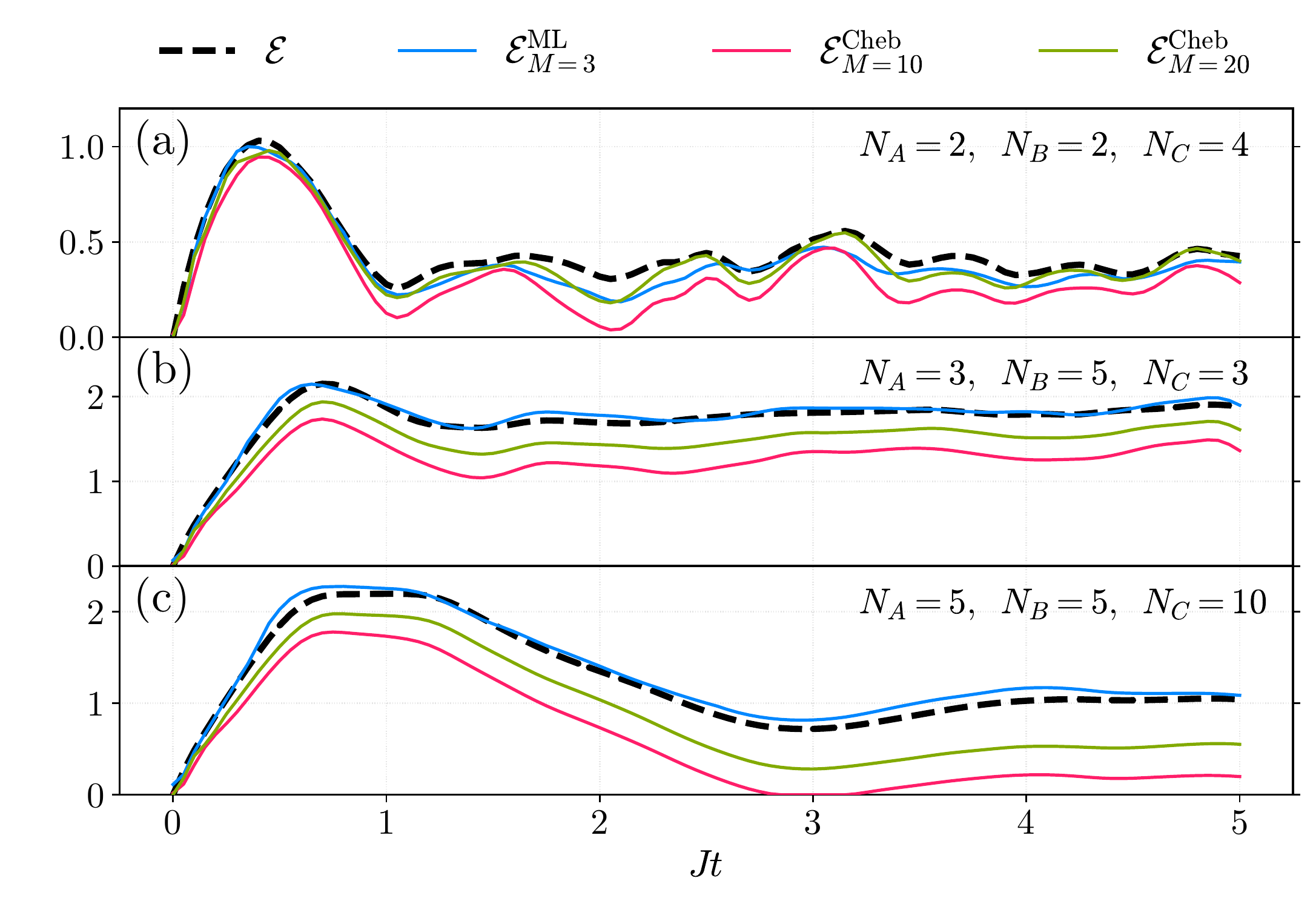}
    \caption{
    \textbf{Estimating entanglement for physical states.}
    Logarithmic negativity, $\mathcal{E}$, and its approximations, using machine learning ($M=3$) and a Chebyshev expansion ($M=10$, $M=20$), as a function of time $J t$ for the quench dynamics of a Heisenberg spin-chain initialised in a Neel-state.
    A variety of system sizes with different partitions is shown here:
    (a) $N=8, N_A=N_B=2$.
    (b) $N=11, N_A=3, N_B=5$;
    (c) $N=20, N_A=N_B=5$.
    }
    \label{fig:evo}
\end{figure}

In Fig.~\ref{fig:evo} we plot the evolution of $\mathcal{E}$ and three approximation methods, as functions of time for three different choices of subsystems.
The three methods are the Chebyshev approximation with $M=10$ and $M=20$, discussed in the appendix, and machine learning with $M=3$, with respective approximate entanglements $\mathcal{E}^\mathrm{Cheb}_{M=10}$, $\mathcal{E}^\mathrm{Cheb}_{M=20}$ and $\mathcal{E}^\mathrm{ML}_{M=3}$.
In Fig.~\ref{fig:evo}(a) we consider a specific partition with $N_A=2$, $N_B=2$ and $N_C=4$. Here, $\mathcal{E}^\mathrm{ML}_{M=3}$ and $\mathcal{E}^\mathrm{Cheb}_{M=20}$ are comparably accurate.
For larger subsystems, as shown in Fig.~\ref{fig:evo}(b) and (c), the machine learning approximation, using only $M=3$ copies, significantly outperforms the Chebyshev approximations, using either $M=10$ or $M=20$ copies.
It is remarkable that despite being trained on a arbitrary set of random states with no knowledge of the underlying physical system, the evolution of $\mathcal{E}$ is accurately captured by the neural network estimator for all partitions and times, with as few as $M=3$ copies.

In the appendix, we explore various other physical situations, including the groundstate of an XX-chain across its phase transition, the fully symmetric W-state, and a quench across the critical point of a transverse Ising chain.

\bigskip

\emph{Conclusions.} ---
The measurement of logarithmic negativity in generic multi-particle mixed states (where Renyi entropies are insufficient to quantify entanglement) has so far relied on the complete reconstruction of a quantum state, which in general requires an exponential number of measurements, and is thus limited to small system sizes.
In this work, we have devised an alternative strategy, based on machine learning, by which we can extract the entanglement from very few measurements.
These measurements are based on two counter-propagating series of swap operators on copies of the state -- techniques for achieving this have already been demonstrated in a number of physical set-ups ranging from quantum dot arrays~\cite{petta2005coherent,schuld2015introduction} to cold atoms in optical lattices~\cite{trotzky2008time,islam2015measuring}.
Our method is based on learning the functional relationship between these measurement outcomes --- the first few moments of the partially transposed density matrix --- and the logarithmic negativity using a neural network.
Remarkably, our method is already very accurate for as few as three copies -- making it very resource efficient and desirable for practical applications -- even for estimating the entanglement of highly entangled physical states, such as those arising in quantum quenches.



JG acknowledges funding from the EPSRC Centre for Doctoral Training in Delivering Quantum Technologies at UCL. SB and AB are supported by the EPSRC grant EP/K004077/1. LB, AB and SB acknowledge financial support by the ERC under Starting Grant 308253 PACOMANEDIA.
LB also acknowledges support from the UK EPSRC grant EP/K034480/1.
AB acknowledges the support from the National Key R\&D Program of China  via the grant No. 2018YFA0306703.

\bibliography{main}

\begin{thebibliography}{83}
\expandafter\ifx\csname natexlab\endcsname\relax\def\natexlab#1{#1}\fi
\expandafter\ifx\csname bibnamefont\endcsname\relax
  \def\bibnamefont#1{#1}\fi
\expandafter\ifx\csname bibfnamefont\endcsname\relax
  \def\bibfnamefont#1{#1}\fi
\expandafter\ifx\csname citenamefont\endcsname\relax
  \def\citenamefont#1{#1}\fi
\expandafter\ifx\csname url\endcsname\relax
  \def\url#1{\texttt{#1}}\fi
\expandafter\ifx\csname urlprefix\endcsname\relax\def\urlprefix{URL }\fi
\providecommand{\bibinfo}[2]{#2}
\providecommand{\eprint}[2][]{\url{#2}}

\bibitem[{\citenamefont{Rao}(1945)}]{rao1945information}
\bibinfo{author}{\bibfnamefont{C.~R.} \bibnamefont{Rao}},
  \bibinfo{journal}{Bull. Calcutta Math. Soc} \textbf{\bibinfo{volume}{37}},
  \bibinfo{pages}{81} (\bibinfo{year}{1945}).

\bibitem[{\citenamefont{Shor}(1999)}]{shor1999polynomial}
\bibinfo{author}{\bibfnamefont{P.~W.} \bibnamefont{Shor}},
  \bibinfo{journal}{SIAM Rev.} \textbf{\bibinfo{volume}{41}},
  \bibinfo{pages}{303} (\bibinfo{year}{1999}).

\bibitem[{\citenamefont{Harrow et~al.}(2009)\citenamefont{Harrow, Hassidim, and
  Lloyd}}]{harrow2009quantum}
\bibinfo{author}{\bibfnamefont{A.~W.} \bibnamefont{Harrow}},
  \bibinfo{author}{\bibfnamefont{A.}~\bibnamefont{Hassidim}}, \bibnamefont{and}
  \bibinfo{author}{\bibfnamefont{S.}~\bibnamefont{Lloyd}},
  \bibinfo{journal}{Phys. Rev. Lett.} \textbf{\bibinfo{volume}{103}},
  \bibinfo{pages}{150502} (\bibinfo{year}{2009}).

\bibitem[{\citenamefont{Bennett et~al.}(1993)\citenamefont{Bennett, Brassard,
  Cr{\'e}peau, Jozsa, Peres, and Wootters}}]{bennett1993teleporting}
\bibinfo{author}{\bibfnamefont{C.~H.} \bibnamefont{Bennett}},
  \bibinfo{author}{\bibfnamefont{G.}~\bibnamefont{Brassard}},
  \bibinfo{author}{\bibfnamefont{C.}~\bibnamefont{Cr{\'e}peau}},
  \bibinfo{author}{\bibfnamefont{R.}~\bibnamefont{Jozsa}},
  \bibinfo{author}{\bibfnamefont{A.}~\bibnamefont{Peres}}, \bibnamefont{and}
  \bibinfo{author}{\bibfnamefont{W.~K.} \bibnamefont{Wootters}},
  \bibinfo{journal}{Phys. Rev. Lett.} \textbf{\bibinfo{volume}{70}},
  \bibinfo{pages}{1895} (\bibinfo{year}{1993}).

\bibitem[{\citenamefont{Bennett and Wiesner}(1992)}]{bennett1992communication}
\bibinfo{author}{\bibfnamefont{C.~H.} \bibnamefont{Bennett}} \bibnamefont{and}
  \bibinfo{author}{\bibfnamefont{S.~J.} \bibnamefont{Wiesner}},
  \bibinfo{journal}{Phys. Rev. Lett.} \textbf{\bibinfo{volume}{69}},
  \bibinfo{pages}{2881} (\bibinfo{year}{1992}).

\bibitem[{\citenamefont{Ekert}(1991)}]{ekert1991quantum}
\bibinfo{author}{\bibfnamefont{A.~K.} \bibnamefont{Ekert}},
  \bibinfo{journal}{Phys. Rev. Lett.} \textbf{\bibinfo{volume}{67}},
  \bibinfo{pages}{661} (\bibinfo{year}{1991}).

\bibitem[{\citenamefont{Amico et~al.}(2008)\citenamefont{Amico, Fazio,
  Osterloh, and Vedral}}]{amico2008entanglement}
\bibinfo{author}{\bibfnamefont{L.}~\bibnamefont{Amico}},
  \bibinfo{author}{\bibfnamefont{R.}~\bibnamefont{Fazio}},
  \bibinfo{author}{\bibfnamefont{A.}~\bibnamefont{Osterloh}}, \bibnamefont{and}
  \bibinfo{author}{\bibfnamefont{V.}~\bibnamefont{Vedral}},
  \bibinfo{journal}{Rev. Mod. Phys.} \textbf{\bibinfo{volume}{80}},
  \bibinfo{pages}{517} (\bibinfo{year}{2008}).

\bibitem[{\citenamefont{Schollw{\"o}ck}(2011)}]{schollwock2011density}
\bibinfo{author}{\bibfnamefont{U.}~\bibnamefont{Schollw{\"o}ck}},
  \bibinfo{journal}{Ann. Phys.} \textbf{\bibinfo{volume}{326}},
  \bibinfo{pages}{96} (\bibinfo{year}{2011}).

\bibitem[{\citenamefont{Horodecki and Ekert}(2002)}]{horodecki2002method}
\bibinfo{author}{\bibfnamefont{P.}~\bibnamefont{Horodecki}} \bibnamefont{and}
  \bibinfo{author}{\bibfnamefont{A.}~\bibnamefont{Ekert}},
  \bibinfo{journal}{Phys. Rev. Lett.} \textbf{\bibinfo{volume}{89}},
  \bibinfo{pages}{127902} (\bibinfo{year}{2002}).

\bibitem[{\citenamefont{Daley et~al.}(2012)\citenamefont{Daley, Pichler,
  Schachenmayer, and Zoller}}]{daley2012measuring}
\bibinfo{author}{\bibfnamefont{A.}~\bibnamefont{Daley}},
  \bibinfo{author}{\bibfnamefont{H.}~\bibnamefont{Pichler}},
  \bibinfo{author}{\bibfnamefont{J.}~\bibnamefont{Schachenmayer}},
  \bibnamefont{and} \bibinfo{author}{\bibfnamefont{P.}~\bibnamefont{Zoller}},
  \bibinfo{journal}{Phys. Rev. Lett.} \textbf{\bibinfo{volume}{109}},
  \bibinfo{pages}{020505} (\bibinfo{year}{2012}).

\bibitem[{\citenamefont{Alves and Jaksch}(2004)}]{alves2004multipartite}
\bibinfo{author}{\bibfnamefont{C.~M.} \bibnamefont{Alves}} \bibnamefont{and}
  \bibinfo{author}{\bibfnamefont{D.}~\bibnamefont{Jaksch}},
  \bibinfo{journal}{Phys. Rev. Lett.} \textbf{\bibinfo{volume}{93}},
  \bibinfo{pages}{110501} (\bibinfo{year}{2004}).

\bibitem[{\citenamefont{Banchi et~al.}(2016{\natexlab{a}})\citenamefont{Banchi,
  Bayat, and Bose}}]{banchi2016entanglement}
\bibinfo{author}{\bibfnamefont{L.}~\bibnamefont{Banchi}},
  \bibinfo{author}{\bibfnamefont{A.}~\bibnamefont{Bayat}}, \bibnamefont{and}
  \bibinfo{author}{\bibfnamefont{S.}~\bibnamefont{Bose}},
  \bibinfo{journal}{Phys. Rev. B} \textbf{\bibinfo{volume}{94}},
  \bibinfo{pages}{241117} (\bibinfo{year}{2016}{\natexlab{a}}).

\bibitem[{\citenamefont{Weedbrook et~al.}(2012)\citenamefont{Weedbrook,
  Pirandola, Garc{\'\i}a-Patr{\'o}n, Cerf, Ralph, Shapiro, and
  Lloyd}}]{weedbrook2012gaussian}
\bibinfo{author}{\bibfnamefont{C.}~\bibnamefont{Weedbrook}},
  \bibinfo{author}{\bibfnamefont{S.}~\bibnamefont{Pirandola}},
  \bibinfo{author}{\bibfnamefont{R.}~\bibnamefont{Garc{\'\i}a-Patr{\'o}n}},
  \bibinfo{author}{\bibfnamefont{N.~J.} \bibnamefont{Cerf}},
  \bibinfo{author}{\bibfnamefont{T.~C.} \bibnamefont{Ralph}},
  \bibinfo{author}{\bibfnamefont{J.~H.} \bibnamefont{Shapiro}},
  \bibnamefont{and} \bibinfo{author}{\bibfnamefont{S.}~\bibnamefont{Lloyd}},
  \bibinfo{journal}{Rev. Mod. Phys.} \textbf{\bibinfo{volume}{84}},
  \bibinfo{pages}{621} (\bibinfo{year}{2012}).

\bibitem[{\citenamefont{Islam et~al.}(2015)\citenamefont{Islam, Ma, Preiss,
  Tai, Lukin, Rispoli, and Greiner}}]{islam2015measuring}
\bibinfo{author}{\bibfnamefont{R.}~\bibnamefont{Islam}},
  \bibinfo{author}{\bibfnamefont{R.}~\bibnamefont{Ma}},
  \bibinfo{author}{\bibfnamefont{P.~M.} \bibnamefont{Preiss}},
  \bibinfo{author}{\bibfnamefont{M.~E.} \bibnamefont{Tai}},
  \bibinfo{author}{\bibfnamefont{A.}~\bibnamefont{Lukin}},
  \bibinfo{author}{\bibfnamefont{M.}~\bibnamefont{Rispoli}}, \bibnamefont{and}
  \bibinfo{author}{\bibfnamefont{M.}~\bibnamefont{Greiner}},
  \bibinfo{journal}{Nature} \textbf{\bibinfo{volume}{528}}, \bibinfo{pages}{77}
  (\bibinfo{year}{2015}).

\bibitem[{\citenamefont{Pitsios et~al.}(2016)\citenamefont{Pitsios, Banchi,
  Rab, Bentivegna, Caprara, Crespi, Spagnolo, Bose, Mataloni, Osellame
  et~al.}}]{pitsios2016photonic}
\bibinfo{author}{\bibfnamefont{I.}~\bibnamefont{Pitsios}},
  \bibinfo{author}{\bibfnamefont{L.}~\bibnamefont{Banchi}},
  \bibinfo{author}{\bibfnamefont{A.~S.} \bibnamefont{Rab}},
  \bibinfo{author}{\bibfnamefont{M.}~\bibnamefont{Bentivegna}},
  \bibinfo{author}{\bibfnamefont{D.}~\bibnamefont{Caprara}},
  \bibinfo{author}{\bibfnamefont{A.}~\bibnamefont{Crespi}},
  \bibinfo{author}{\bibfnamefont{N.}~\bibnamefont{Spagnolo}},
  \bibinfo{author}{\bibfnamefont{S.}~\bibnamefont{Bose}},
  \bibinfo{author}{\bibfnamefont{P.}~\bibnamefont{Mataloni}},
  \bibinfo{author}{\bibfnamefont{R.}~\bibnamefont{Osellame}},
  \bibnamefont{et~al.}, \bibinfo{journal}{arXiv:1603.02669}
  (\bibinfo{year}{2016}).

\bibitem[{\citenamefont{Reznik}(2003)}]{reznik2003entanglement}
\bibinfo{author}{\bibfnamefont{B.}~\bibnamefont{Reznik}},
  \bibinfo{journal}{Found. Phys.} \textbf{\bibinfo{volume}{33}},
  \bibinfo{pages}{167} (\bibinfo{year}{2003}).

\bibitem[{\citenamefont{Bayat et~al.}(2010)\citenamefont{Bayat, Sodano, and
  Bose}}]{bayat2010negativity}
\bibinfo{author}{\bibfnamefont{A.}~\bibnamefont{Bayat}},
  \bibinfo{author}{\bibfnamefont{P.}~\bibnamefont{Sodano}}, \bibnamefont{and}
  \bibinfo{author}{\bibfnamefont{S.}~\bibnamefont{Bose}},
  \bibinfo{journal}{Phys. Rev. B} \textbf{\bibinfo{volume}{81}},
  \bibinfo{pages}{064429} (\bibinfo{year}{2010}).

\bibitem[{\citenamefont{Alkurtass et~al.}(2016)\citenamefont{Alkurtass, Bayat,
  Affleck, Bose, Johannesson, Sodano, S{\o}rensen, and
  Le~Hur}}]{alkurtass2016entanglement}
\bibinfo{author}{\bibfnamefont{B.}~\bibnamefont{Alkurtass}},
  \bibinfo{author}{\bibfnamefont{A.}~\bibnamefont{Bayat}},
  \bibinfo{author}{\bibfnamefont{I.}~\bibnamefont{Affleck}},
  \bibinfo{author}{\bibfnamefont{S.}~\bibnamefont{Bose}},
  \bibinfo{author}{\bibfnamefont{H.}~\bibnamefont{Johannesson}},
  \bibinfo{author}{\bibfnamefont{P.}~\bibnamefont{Sodano}},
  \bibinfo{author}{\bibfnamefont{E.~S.} \bibnamefont{S{\o}rensen}},
  \bibnamefont{and} \bibinfo{author}{\bibfnamefont{K.}~\bibnamefont{Le~Hur}},
  \bibinfo{journal}{Phys. Rev. B} \textbf{\bibinfo{volume}{93}},
  \bibinfo{pages}{081106} (\bibinfo{year}{2016}).

\bibitem[{\citenamefont{Gray et~al.}(2017)\citenamefont{Gray, Bose, and
  Bayat}}]{gray2017many}
\bibinfo{author}{\bibfnamefont{J.}~\bibnamefont{Gray}},
  \bibinfo{author}{\bibfnamefont{S.}~\bibnamefont{Bose}}, \bibnamefont{and}
  \bibinfo{author}{\bibfnamefont{A.}~\bibnamefont{Bayat}},
  \bibinfo{journal}{arXiv:1704.00738}  (\bibinfo{year}{2017}).

\bibitem[{\citenamefont{Banchi et~al.}(2009)\citenamefont{Banchi, Colomo, and
  Verrucchi}}]{banchi2009finite}
\bibinfo{author}{\bibfnamefont{L.}~\bibnamefont{Banchi}},
  \bibinfo{author}{\bibfnamefont{F.}~\bibnamefont{Colomo}}, \bibnamefont{and}
  \bibinfo{author}{\bibfnamefont{P.}~\bibnamefont{Verrucchi}},
  \bibinfo{journal}{Phys. Rev. A} \textbf{\bibinfo{volume}{80}},
  \bibinfo{pages}{022341} (\bibinfo{year}{2009}).

\bibitem[{\citenamefont{Roscilde et~al.}(2004)\citenamefont{Roscilde,
  Verrucchi, Fubini, Haas, and Tognetti}}]{roscilde2004studying}
\bibinfo{author}{\bibfnamefont{T.}~\bibnamefont{Roscilde}},
  \bibinfo{author}{\bibfnamefont{P.}~\bibnamefont{Verrucchi}},
  \bibinfo{author}{\bibfnamefont{A.}~\bibnamefont{Fubini}},
  \bibinfo{author}{\bibfnamefont{S.}~\bibnamefont{Haas}}, \bibnamefont{and}
  \bibinfo{author}{\bibfnamefont{V.}~\bibnamefont{Tognetti}},
  \bibinfo{journal}{Phys. Rev. Lett.} \textbf{\bibinfo{volume}{93}},
  \bibinfo{pages}{167203} (\bibinfo{year}{2004}).

\bibitem[{\citenamefont{Horodecki et~al.}(1996)\citenamefont{Horodecki,
  Horodecki, and Horodecki}}]{HORODECKI19961}
\bibinfo{author}{\bibfnamefont{M.}~\bibnamefont{Horodecki}},
  \bibinfo{author}{\bibfnamefont{P.}~\bibnamefont{Horodecki}},
  \bibnamefont{and}
  \bibinfo{author}{\bibfnamefont{R.}~\bibnamefont{Horodecki}},
  \bibinfo{journal}{Phys. Lett. A} \textbf{\bibinfo{volume}{223}},
  \bibinfo{pages}{1} (\bibinfo{year}{1996}).

\bibitem[{\citenamefont{Terhal}(2000)}]{terhal2000bell}
\bibinfo{author}{\bibfnamefont{B.~M.} \bibnamefont{Terhal}},
  \bibinfo{journal}{Phys. Lett. A} \textbf{\bibinfo{volume}{271}},
  \bibinfo{pages}{319} (\bibinfo{year}{2000}).

\bibitem[{\citenamefont{Lewenstein et~al.}(2000)\citenamefont{Lewenstein,
  Kraus, Cirac, and Horodecki}}]{lewenstein2000optimization}
\bibinfo{author}{\bibfnamefont{M.}~\bibnamefont{Lewenstein}},
  \bibinfo{author}{\bibfnamefont{B.}~\bibnamefont{Kraus}},
  \bibinfo{author}{\bibfnamefont{J.}~\bibnamefont{Cirac}}, \bibnamefont{and}
  \bibinfo{author}{\bibfnamefont{P.}~\bibnamefont{Horodecki}},
  \bibinfo{journal}{Phys. Rev. A} \textbf{\bibinfo{volume}{62}},
  \bibinfo{pages}{052310} (\bibinfo{year}{2000}).

\bibitem[{\citenamefont{Brandao}(2005)}]{brandao2005quantifying}
\bibinfo{author}{\bibfnamefont{F.~G.} \bibnamefont{Brandao}},
  \bibinfo{journal}{Physical Review A} \textbf{\bibinfo{volume}{72}},
  \bibinfo{pages}{022310} (\bibinfo{year}{2005}).

\bibitem[{\citenamefont{Audenaert and
  Plenio}(2006)}]{audenaert2006correlations}
\bibinfo{author}{\bibfnamefont{K.}~\bibnamefont{Audenaert}} \bibnamefont{and}
  \bibinfo{author}{\bibfnamefont{M.}~\bibnamefont{Plenio}},
  \bibinfo{journal}{New Journal of Physics} \textbf{\bibinfo{volume}{8}},
  \bibinfo{pages}{266} (\bibinfo{year}{2006}).

\bibitem[{\citenamefont{Cavalcanti and
  Terra~Cunha}(2006)}]{cavalcanti2006estimating}
\bibinfo{author}{\bibfnamefont{D.}~\bibnamefont{Cavalcanti}} \bibnamefont{and}
  \bibinfo{author}{\bibfnamefont{M.~O.} \bibnamefont{Terra~Cunha}},
  \bibinfo{journal}{Applied physics letters} \textbf{\bibinfo{volume}{89}},
  \bibinfo{pages}{084102} (\bibinfo{year}{2006}).

\bibitem[{\citenamefont{Eisert et~al.}(2007)\citenamefont{Eisert, Brandao, and
  Audenaert}}]{eisert2007quantitative}
\bibinfo{author}{\bibfnamefont{J.}~\bibnamefont{Eisert}},
  \bibinfo{author}{\bibfnamefont{F.~G.} \bibnamefont{Brandao}},
  \bibnamefont{and} \bibinfo{author}{\bibfnamefont{K.~M.}
  \bibnamefont{Audenaert}}, \bibinfo{journal}{New Journal of Physics}
  \textbf{\bibinfo{volume}{9}}, \bibinfo{pages}{46} (\bibinfo{year}{2007}).

\bibitem[{\citenamefont{G{\"u}hne et~al.}(2007)\citenamefont{G{\"u}hne,
  Reimpell, and Werner}}]{guhne2007estimating}
\bibinfo{author}{\bibfnamefont{O.}~\bibnamefont{G{\"u}hne}},
  \bibinfo{author}{\bibfnamefont{M.}~\bibnamefont{Reimpell}}, \bibnamefont{and}
  \bibinfo{author}{\bibfnamefont{R.}~\bibnamefont{Werner}},
  \bibinfo{journal}{Physical review letters} \textbf{\bibinfo{volume}{98}},
  \bibinfo{pages}{110502} (\bibinfo{year}{2007}).

\bibitem[{\citenamefont{G{\"u}hne et~al.}(2008)\citenamefont{G{\"u}hne,
  Reimpell, and Werner}}]{guhne2008lower}
\bibinfo{author}{\bibfnamefont{O.}~\bibnamefont{G{\"u}hne}},
  \bibinfo{author}{\bibfnamefont{M.}~\bibnamefont{Reimpell}}, \bibnamefont{and}
  \bibinfo{author}{\bibfnamefont{R.}~\bibnamefont{Werner}},
  \bibinfo{journal}{Physical Review A} \textbf{\bibinfo{volume}{77}},
  \bibinfo{pages}{052317} (\bibinfo{year}{2008}).

\bibitem[{\citenamefont{Carteret}(2016)}]{carteret2016estimating}
\bibinfo{author}{\bibfnamefont{H.~A.} \bibnamefont{Carteret}},
  \bibinfo{journal}{arXiv preprint arXiv:1605.08751}  (\bibinfo{year}{2016}).

\bibitem[{\citenamefont{Horodecki et~al.}(2009)\citenamefont{Horodecki,
  Horodecki, Horodecki, and Horodecki}}]{horodecki2009quantum}
\bibinfo{author}{\bibfnamefont{R.}~\bibnamefont{Horodecki}},
  \bibinfo{author}{\bibfnamefont{P.}~\bibnamefont{Horodecki}},
  \bibinfo{author}{\bibfnamefont{M.}~\bibnamefont{Horodecki}},
  \bibnamefont{and}
  \bibinfo{author}{\bibfnamefont{K.}~\bibnamefont{Horodecki}},
  \bibinfo{journal}{Rev. Mod. Phys.} \textbf{\bibinfo{volume}{81}},
  \bibinfo{pages}{865} (\bibinfo{year}{2009}).

\bibitem[{\citenamefont{Plenio and Virmani}(2007)}]{plenio2007introduction}
\bibinfo{author}{\bibfnamefont{M.~B.} \bibnamefont{Plenio}} \bibnamefont{and}
  \bibinfo{author}{\bibfnamefont{S.}~\bibnamefont{Virmani}},
  \bibinfo{journal}{Quantum Inf. Comput.} \textbf{\bibinfo{volume}{7}},
  \bibinfo{pages}{1} (\bibinfo{year}{2007}).

\bibitem[{\citenamefont{Wootters}(1998)}]{wootters1998entanglement}
\bibinfo{author}{\bibfnamefont{W.~K.} \bibnamefont{Wootters}},
  \bibinfo{journal}{Phys. Rev. Lett.} \textbf{\bibinfo{volume}{80}},
  \bibinfo{pages}{2245} (\bibinfo{year}{1998}).

\bibitem[{\citenamefont{Wolf et~al.}(2004)\citenamefont{Wolf, Giedke,
  Kr{\"u}ger, Werner, and Cirac}}]{wolf2004gaussian}
\bibinfo{author}{\bibfnamefont{M.~M.} \bibnamefont{Wolf}},
  \bibinfo{author}{\bibfnamefont{G.}~\bibnamefont{Giedke}},
  \bibinfo{author}{\bibfnamefont{O.}~\bibnamefont{Kr{\"u}ger}},
  \bibinfo{author}{\bibfnamefont{R.}~\bibnamefont{Werner}}, \bibnamefont{and}
  \bibinfo{author}{\bibfnamefont{J.~I.} \bibnamefont{Cirac}},
  \bibinfo{journal}{Phys. Rev. A} \textbf{\bibinfo{volume}{69}},
  \bibinfo{pages}{052320} (\bibinfo{year}{2004}).

\bibitem[{\citenamefont{{\.Z}yczkowski
  et~al.}(1998)\citenamefont{{\.Z}yczkowski, Horodecki, Sanpera, and
  Lewenstein}}]{zyczkowski1998volume}
\bibinfo{author}{\bibfnamefont{K.}~\bibnamefont{{\.Z}yczkowski}},
  \bibinfo{author}{\bibfnamefont{P.}~\bibnamefont{Horodecki}},
  \bibinfo{author}{\bibfnamefont{A.}~\bibnamefont{Sanpera}}, \bibnamefont{and}
  \bibinfo{author}{\bibfnamefont{M.}~\bibnamefont{Lewenstein}},
  \bibinfo{journal}{Phys. Rev. A} \textbf{\bibinfo{volume}{58}},
  \bibinfo{pages}{883} (\bibinfo{year}{1998}).

\bibitem[{\citenamefont{Lee et~al.}(2000)\citenamefont{Lee, Kim, Park, and
  Lee}}]{lee2000partial}
\bibinfo{author}{\bibfnamefont{J.}~\bibnamefont{Lee}},
  \bibinfo{author}{\bibfnamefont{M.}~\bibnamefont{Kim}},
  \bibinfo{author}{\bibfnamefont{Y.}~\bibnamefont{Park}}, \bibnamefont{and}
  \bibinfo{author}{\bibfnamefont{S.}~\bibnamefont{Lee}}, \bibinfo{journal}{J.
  Mod. Opt.} \textbf{\bibinfo{volume}{47}}, \bibinfo{pages}{2151}
  (\bibinfo{year}{2000}).

\bibitem[{\citenamefont{Vidal and Werner}(2002)}]{vidal2002computable}
\bibinfo{author}{\bibfnamefont{G.}~\bibnamefont{Vidal}} \bibnamefont{and}
  \bibinfo{author}{\bibfnamefont{R.~F.} \bibnamefont{Werner}},
  \bibinfo{journal}{Phys. Rev. A} \textbf{\bibinfo{volume}{65}},
  \bibinfo{pages}{032314} (\bibinfo{year}{2002}).

\bibitem[{\citenamefont{Plenio}(2005)}]{plenio2005logarithmic}
\bibinfo{author}{\bibfnamefont{M.~B.} \bibnamefont{Plenio}},
  \bibinfo{journal}{Phys. Rev. Lett.} \textbf{\bibinfo{volume}{95}},
  \bibinfo{pages}{090503} (\bibinfo{year}{2005}).

\bibitem[{\citenamefont{Huang}(2014)}]{huang2014computing}
\bibinfo{author}{\bibfnamefont{Y.}~\bibnamefont{Huang}}, \bibinfo{journal}{New
  journal of physics} \textbf{\bibinfo{volume}{16}}, \bibinfo{pages}{033027}
  (\bibinfo{year}{2014}).

\bibitem[{\citenamefont{Lanyon et~al.}(2016)\citenamefont{Lanyon, Maier,
  Holz{\"a}pfel, Baumgratz, Hempel, Jurcevic, Dhand, Buyskikh, Daley, Cramer
  et~al.}}]{lanyon2016efficient}
\bibinfo{author}{\bibfnamefont{B.}~\bibnamefont{Lanyon}},
  \bibinfo{author}{\bibfnamefont{C.}~\bibnamefont{Maier}},
  \bibinfo{author}{\bibfnamefont{M.}~\bibnamefont{Holz{\"a}pfel}},
  \bibinfo{author}{\bibfnamefont{T.}~\bibnamefont{Baumgratz}},
  \bibinfo{author}{\bibfnamefont{C.}~\bibnamefont{Hempel}},
  \bibinfo{author}{\bibfnamefont{P.}~\bibnamefont{Jurcevic}},
  \bibinfo{author}{\bibfnamefont{I.}~\bibnamefont{Dhand}},
  \bibinfo{author}{\bibfnamefont{A.}~\bibnamefont{Buyskikh}},
  \bibinfo{author}{\bibfnamefont{A.}~\bibnamefont{Daley}},
  \bibinfo{author}{\bibfnamefont{M.}~\bibnamefont{Cramer}},
  \bibnamefont{et~al.}, \bibinfo{journal}{arXiv:1612.08000}
  (\bibinfo{year}{2016}).

\bibitem[{\citenamefont{Calabrese et~al.}(2012)\citenamefont{Calabrese, Cardy,
  and Tonni}}]{calabrese2012entanglement}
\bibinfo{author}{\bibfnamefont{P.}~\bibnamefont{Calabrese}},
  \bibinfo{author}{\bibfnamefont{J.}~\bibnamefont{Cardy}}, \bibnamefont{and}
  \bibinfo{author}{\bibfnamefont{E.}~\bibnamefont{Tonni}},
  \bibinfo{journal}{Phys. Rev. Lett.} \textbf{\bibinfo{volume}{109}},
  \bibinfo{pages}{130502} (\bibinfo{year}{2012}).

\bibitem[{\citenamefont{D'Ariano et~al.}(2003)\citenamefont{D'Ariano, Paris,
  and Sacchi}}]{d2003quantum}
\bibinfo{author}{\bibfnamefont{G.~M.} \bibnamefont{D'Ariano}},
  \bibinfo{author}{\bibfnamefont{M.~G.} \bibnamefont{Paris}}, \bibnamefont{and}
  \bibinfo{author}{\bibfnamefont{M.~F.} \bibnamefont{Sacchi}},
  \bibinfo{journal}{Adv. Imag. Electr. Phys.} \textbf{\bibinfo{volume}{128}},
  \bibinfo{pages}{206} (\bibinfo{year}{2003}).

\bibitem[{\citenamefont{Cramer et~al.}(2010)\citenamefont{Cramer, Plenio,
  Flammia, Somma, Gross, Bartlett, Landon-Cardinal, Poulin, and
  Liu}}]{cramer2010efficient}
\bibinfo{author}{\bibfnamefont{M.}~\bibnamefont{Cramer}},
  \bibinfo{author}{\bibfnamefont{M.~B.} \bibnamefont{Plenio}},
  \bibinfo{author}{\bibfnamefont{S.~T.} \bibnamefont{Flammia}},
  \bibinfo{author}{\bibfnamefont{R.}~\bibnamefont{Somma}},
  \bibinfo{author}{\bibfnamefont{D.}~\bibnamefont{Gross}},
  \bibinfo{author}{\bibfnamefont{S.~D.} \bibnamefont{Bartlett}},
  \bibinfo{author}{\bibfnamefont{O.}~\bibnamefont{Landon-Cardinal}},
  \bibinfo{author}{\bibfnamefont{D.}~\bibnamefont{Poulin}}, \bibnamefont{and}
  \bibinfo{author}{\bibfnamefont{Y.-K.} \bibnamefont{Liu}},
  \bibinfo{journal}{Nat. Commun.} \textbf{\bibinfo{volume}{1}},
  \bibinfo{pages}{149} (\bibinfo{year}{2010}).

\bibitem[{\citenamefont{Torlai et~al.}(2018)\citenamefont{Torlai, Mazzola,
  Carrasquilla, Troyer, Melko, and Carleo}}]{torlai2018neural}
\bibinfo{author}{\bibfnamefont{G.}~\bibnamefont{Torlai}},
  \bibinfo{author}{\bibfnamefont{G.}~\bibnamefont{Mazzola}},
  \bibinfo{author}{\bibfnamefont{J.}~\bibnamefont{Carrasquilla}},
  \bibinfo{author}{\bibfnamefont{M.}~\bibnamefont{Troyer}},
  \bibinfo{author}{\bibfnamefont{R.}~\bibnamefont{Melko}}, \bibnamefont{and}
  \bibinfo{author}{\bibfnamefont{G.}~\bibnamefont{Carleo}},
  \bibinfo{journal}{Nat. Phys.} \textbf{\bibinfo{volume}{14}},
  \bibinfo{pages}{447} (\bibinfo{year}{2018}).

\bibitem[{\citenamefont{Torlai and Melko}(2018)}]{torlai2018latent}
\bibinfo{author}{\bibfnamefont{G.}~\bibnamefont{Torlai}} \bibnamefont{and}
  \bibinfo{author}{\bibfnamefont{R.~G.} \bibnamefont{Melko}},
  \bibinfo{journal}{Phys. Rev. Lett.} \textbf{\bibinfo{volume}{120}},
  \bibinfo{pages}{240503} (\bibinfo{year}{2018}).

\bibitem[{\citenamefont{Bengtsson and
  {\.Z}yczkowski}(2006)}]{bengtsson2006geometry}
\bibinfo{author}{\bibfnamefont{I.}~\bibnamefont{Bengtsson}} \bibnamefont{and}
  \bibinfo{author}{\bibfnamefont{K.}~\bibnamefont{{\.Z}yczkowski}},
  \emph{\bibinfo{title}{Geometry of quantum states: an introduction to quantum
  entanglement}} (\bibinfo{publisher}{Cambridge Univ Pr},
  \bibinfo{year}{2006}).

\bibitem[{asy()}]{asymptoticContinuityNote}
\bibinfo{note}{See the appendix section `Comparison with approximate state
  reconstruction methods'.}

\bibitem[{\citenamefont{Carteret}(2005)}]{carteret2005noiseless}
\bibinfo{author}{\bibfnamefont{H.~A.} \bibnamefont{Carteret}},
  \bibinfo{journal}{Phys. Rev. Lett.} \textbf{\bibinfo{volume}{94}},
  \bibinfo{pages}{040502} (\bibinfo{year}{2005}).

\bibitem[{\citenamefont{Cai and Song}(2008)}]{cai2008novel}
\bibinfo{author}{\bibfnamefont{J.}~\bibnamefont{Cai}} \bibnamefont{and}
  \bibinfo{author}{\bibfnamefont{W.}~\bibnamefont{Song}},
  \bibinfo{journal}{Phys. Rev. Lett.} \textbf{\bibinfo{volume}{101}},
  \bibinfo{pages}{190503} (\bibinfo{year}{2008}).

\bibitem[{\citenamefont{Bartkiewicz et~al.}(2015)\citenamefont{Bartkiewicz,
  Horodecki, Lemr, Miranowicz, and \ifmmode~\dot{Z}\else
  \.{Z}\fi{}yczkowski}}]{Bartkiewicz2014method}
\bibinfo{author}{\bibfnamefont{K.}~\bibnamefont{Bartkiewicz}},
  \bibinfo{author}{\bibfnamefont{P.}~\bibnamefont{Horodecki}},
  \bibinfo{author}{\bibfnamefont{K.}~\bibnamefont{Lemr}},
  \bibinfo{author}{\bibfnamefont{A.}~\bibnamefont{Miranowicz}},
  \bibnamefont{and}
  \bibinfo{author}{\bibfnamefont{K.}~\bibnamefont{\ifmmode~\dot{Z}\else
  \.{Z}\fi{}yczkowski}}, \bibinfo{journal}{Phys. Rev. A}
  \textbf{\bibinfo{volume}{91}}, \bibinfo{pages}{032315}
  (\bibinfo{year}{2015}).

\bibitem[{\citenamefont{Carrasquilla and
  Melko}(2017)}]{carrasquilla2017machine}
\bibinfo{author}{\bibfnamefont{J.}~\bibnamefont{Carrasquilla}}
  \bibnamefont{and} \bibinfo{author}{\bibfnamefont{R.~G.} \bibnamefont{Melko}},
  \bibinfo{journal}{Nat. Phys.} \textbf{\bibinfo{volume}{13}},
  \bibinfo{pages}{431} (\bibinfo{year}{2017}).

\bibitem[{\citenamefont{Carleo and Troyer}(2017)}]{carleo2017solving}
\bibinfo{author}{\bibfnamefont{G.}~\bibnamefont{Carleo}} \bibnamefont{and}
  \bibinfo{author}{\bibfnamefont{M.}~\bibnamefont{Troyer}},
  \bibinfo{journal}{Science} \textbf{\bibinfo{volume}{355}},
  \bibinfo{pages}{602} (\bibinfo{year}{2017}).

\bibitem[{\citenamefont{Hentschel and Sanders}(2010)}]{hentschel2010machine}
\bibinfo{author}{\bibfnamefont{A.}~\bibnamefont{Hentschel}} \bibnamefont{and}
  \bibinfo{author}{\bibfnamefont{B.~C.} \bibnamefont{Sanders}},
  \bibinfo{journal}{Phys. Rev. Lett.} \textbf{\bibinfo{volume}{104}},
  \bibinfo{pages}{063603} (\bibinfo{year}{2010}).

\bibitem[{\citenamefont{Banchi et~al.}(2016{\natexlab{b}})\citenamefont{Banchi,
  Pancotti, and Bose}}]{banchi2016quantum}
\bibinfo{author}{\bibfnamefont{L.}~\bibnamefont{Banchi}},
  \bibinfo{author}{\bibfnamefont{N.}~\bibnamefont{Pancotti}}, \bibnamefont{and}
  \bibinfo{author}{\bibfnamefont{S.}~\bibnamefont{Bose}}, \bibinfo{journal}{npj
  Quantum Inf.} \textbf{\bibinfo{volume}{2}}, \bibinfo{pages}{16019}
  (\bibinfo{year}{2016}{\natexlab{b}}).

\bibitem[{\citenamefont{Petta et~al.}(2005)\citenamefont{Petta, Johnson,
  Taylor, Laird, Yacoby, Lukin, Marcus, Hanson, and
  Gossard}}]{petta2005coherent}
\bibinfo{author}{\bibfnamefont{J.~R.} \bibnamefont{Petta}},
  \bibinfo{author}{\bibfnamefont{A.~C.} \bibnamefont{Johnson}},
  \bibinfo{author}{\bibfnamefont{J.~M.} \bibnamefont{Taylor}},
  \bibinfo{author}{\bibfnamefont{E.~A.} \bibnamefont{Laird}},
  \bibinfo{author}{\bibfnamefont{A.}~\bibnamefont{Yacoby}},
  \bibinfo{author}{\bibfnamefont{M.~D.} \bibnamefont{Lukin}},
  \bibinfo{author}{\bibfnamefont{C.~M.} \bibnamefont{Marcus}},
  \bibinfo{author}{\bibfnamefont{M.~P.} \bibnamefont{Hanson}},
  \bibnamefont{and} \bibinfo{author}{\bibfnamefont{A.~C.}
  \bibnamefont{Gossard}}, \bibinfo{journal}{Science}
  \textbf{\bibinfo{volume}{309}}, \bibinfo{pages}{2180} (\bibinfo{year}{2005}).

\bibitem[{\citenamefont{Viano}(1991)}]{viano1991solution}
\bibinfo{author}{\bibfnamefont{G.}~\bibnamefont{Viano}}, \bibinfo{journal}{J.
  Math. Anal. Appl.} \textbf{\bibinfo{volume}{156}}, \bibinfo{pages}{410}
  (\bibinfo{year}{1991}).

\bibitem[{\citenamefont{Rana}(2013)}]{rana2013negative}
\bibinfo{author}{\bibfnamefont{S.}~\bibnamefont{Rana}}, \bibinfo{journal}{Phys.
  Rev. A} \textbf{\bibinfo{volume}{87}}, \bibinfo{pages}{054301}
  (\bibinfo{year}{2013}).

\bibitem[{\citenamefont{Ekert et~al.}(2002)\citenamefont{Ekert, Alves, Oi,
  Horodecki, Horodecki, and Kwek}}]{ekert2002direct}
\bibinfo{author}{\bibfnamefont{A.~K.} \bibnamefont{Ekert}},
  \bibinfo{author}{\bibfnamefont{C.~M.} \bibnamefont{Alves}},
  \bibinfo{author}{\bibfnamefont{D.~K.} \bibnamefont{Oi}},
  \bibinfo{author}{\bibfnamefont{M.}~\bibnamefont{Horodecki}},
  \bibinfo{author}{\bibfnamefont{P.}~\bibnamefont{Horodecki}},
  \bibnamefont{and} \bibinfo{author}{\bibfnamefont{L.~C.} \bibnamefont{Kwek}},
  \bibinfo{journal}{Phys. Rev. Lett.} \textbf{\bibinfo{volume}{88}},
  \bibinfo{pages}{217901} (\bibinfo{year}{2002}).

\bibitem[{\citenamefont{Reck et~al.}(1994)\citenamefont{Reck, Zeilinger,
  Bernstein, and Bertani}}]{reck1994experimental}
\bibinfo{author}{\bibfnamefont{M.}~\bibnamefont{Reck}},
  \bibinfo{author}{\bibfnamefont{A.}~\bibnamefont{Zeilinger}},
  \bibinfo{author}{\bibfnamefont{H.~J.} \bibnamefont{Bernstein}},
  \bibnamefont{and} \bibinfo{author}{\bibfnamefont{P.}~\bibnamefont{Bertani}},
  \bibinfo{journal}{Phys. Rev. Lett.} \textbf{\bibinfo{volume}{73}},
  \bibinfo{pages}{58} (\bibinfo{year}{1994}).

\bibitem[{\citenamefont{Abanin and Demler}(2012)}]{abanin2012measuring}
\bibinfo{author}{\bibfnamefont{D.~A.} \bibnamefont{Abanin}} \bibnamefont{and}
  \bibinfo{author}{\bibfnamefont{E.}~\bibnamefont{Demler}},
  \bibinfo{journal}{Phys. Rev. Lett.} \textbf{\bibinfo{volume}{109}},
  \bibinfo{pages}{020504} (\bibinfo{year}{2012}).

\bibitem[{\citenamefont{De~Nobili et~al.}(2015)\citenamefont{De~Nobili, Coser,
  and Tonni}}]{de2015entanglement}
\bibinfo{author}{\bibfnamefont{C.}~\bibnamefont{De~Nobili}},
  \bibinfo{author}{\bibfnamefont{A.}~\bibnamefont{Coser}}, \bibnamefont{and}
  \bibinfo{author}{\bibfnamefont{E.}~\bibnamefont{Tonni}}, \bibinfo{journal}{J.
  Stat. Mech: Theory Exp.} \textbf{\bibinfo{volume}{2015}},
  \bibinfo{pages}{P06021} (\bibinfo{year}{2015}).

\bibitem[{\citenamefont{Cristianini and
  Shawe-Taylor}(2000)}]{cristianini2000introduction}
\bibinfo{author}{\bibfnamefont{N.}~\bibnamefont{Cristianini}} \bibnamefont{and}
  \bibinfo{author}{\bibfnamefont{J.}~\bibnamefont{Shawe-Taylor}},
  \emph{\bibinfo{title}{An introduction to support vector machines and other
  kernel-based learning methods}} (\bibinfo{publisher}{Cambridge university
  press}, \bibinfo{year}{2000}).

\bibitem[{\citenamefont{Ho}(1998)}]{ho1998random}
\bibinfo{author}{\bibfnamefont{T.~K.} \bibnamefont{Ho}}, \bibinfo{journal}{IEEE
  Trans. Pattern Anal. Mach. Intell.} \textbf{\bibinfo{volume}{20}},
  \bibinfo{pages}{832} (\bibinfo{year}{1998}).

\bibitem[{\citenamefont{Rojas}(2013)}]{rojas2013neural}
\bibinfo{author}{\bibfnamefont{R.}~\bibnamefont{Rojas}},
  \emph{\bibinfo{title}{Neural networks: a systematic introduction}}
  (\bibinfo{publisher}{Springer Science \& Business Media},
  \bibinfo{year}{2013}).

\bibitem[{\citenamefont{Schmidhuber}(2015)}]{schmidhuber2015deep}
\bibinfo{author}{\bibfnamefont{J.}~\bibnamefont{Schmidhuber}},
  \bibinfo{journal}{Neural Netw.} \textbf{\bibinfo{volume}{61}},
  \bibinfo{pages}{85} (\bibinfo{year}{2015}).

\bibitem[{\citenamefont{Barmettler et~al.}(2009)\citenamefont{Barmettler, Punk,
  Gritsev, Demler, and Altman}}]{barmettler2009relaxation}
\bibinfo{author}{\bibfnamefont{P.}~\bibnamefont{Barmettler}},
  \bibinfo{author}{\bibfnamefont{M.}~\bibnamefont{Punk}},
  \bibinfo{author}{\bibfnamefont{V.}~\bibnamefont{Gritsev}},
  \bibinfo{author}{\bibfnamefont{E.}~\bibnamefont{Demler}}, \bibnamefont{and}
  \bibinfo{author}{\bibfnamefont{E.}~\bibnamefont{Altman}},
  \bibinfo{journal}{Phys. Rev. Lett.} \textbf{\bibinfo{volume}{102}},
  \bibinfo{pages}{130603} (\bibinfo{year}{2009}).

\bibitem[{\citenamefont{Nanduri et~al.}(2014)\citenamefont{Nanduri, Kim, and
  Huse}}]{nanduri2014entanglement}
\bibinfo{author}{\bibfnamefont{A.}~\bibnamefont{Nanduri}},
  \bibinfo{author}{\bibfnamefont{H.}~\bibnamefont{Kim}}, \bibnamefont{and}
  \bibinfo{author}{\bibfnamefont{D.~A.} \bibnamefont{Huse}},
  \bibinfo{journal}{Phys. Rev. B} \textbf{\bibinfo{volume}{90}},
  \bibinfo{pages}{064201} (\bibinfo{year}{2014}).

\bibitem[{\citenamefont{Popescu et~al.}(2006)\citenamefont{Popescu, Short, and
  Winter}}]{popescu2006entanglement}
\bibinfo{author}{\bibfnamefont{S.}~\bibnamefont{Popescu}},
  \bibinfo{author}{\bibfnamefont{A.~J.} \bibnamefont{Short}}, \bibnamefont{and}
  \bibinfo{author}{\bibfnamefont{A.}~\bibnamefont{Winter}},
  \bibinfo{journal}{Nat. Phys.} \textbf{\bibinfo{volume}{2}},
  \bibinfo{pages}{754} (\bibinfo{year}{2006}).

\bibitem[{\citenamefont{Hamma et~al.}(2012)\citenamefont{Hamma, Santra, and
  Zanardi}}]{hamma2012quantum}
\bibinfo{author}{\bibfnamefont{A.}~\bibnamefont{Hamma}},
  \bibinfo{author}{\bibfnamefont{S.}~\bibnamefont{Santra}}, \bibnamefont{and}
  \bibinfo{author}{\bibfnamefont{P.}~\bibnamefont{Zanardi}},
  \bibinfo{journal}{Phys. Rev. Lett.} \textbf{\bibinfo{volume}{109}},
  \bibinfo{pages}{040502} (\bibinfo{year}{2012}).

\bibitem[{\citenamefont{Schuld et~al.}(2015)\citenamefont{Schuld, Sinayskiy,
  and Petruccione}}]{schuld2015introduction}
\bibinfo{author}{\bibfnamefont{M.}~\bibnamefont{Schuld}},
  \bibinfo{author}{\bibfnamefont{I.}~\bibnamefont{Sinayskiy}},
  \bibnamefont{and}
  \bibinfo{author}{\bibfnamefont{F.}~\bibnamefont{Petruccione}},
  \bibinfo{journal}{Contemp. Phys.} \textbf{\bibinfo{volume}{56}},
  \bibinfo{pages}{172} (\bibinfo{year}{2015}).

\bibitem[{\citenamefont{Trotzky et~al.}(2008)\citenamefont{Trotzky, Cheinet,
  F{\"o}lling, Feld, Schnorrberger, Rey, Polkovnikov, Demler, Lukin, and
  Bloch}}]{trotzky2008time}
\bibinfo{author}{\bibfnamefont{S.}~\bibnamefont{Trotzky}},
  \bibinfo{author}{\bibfnamefont{P.}~\bibnamefont{Cheinet}},
  \bibinfo{author}{\bibfnamefont{S.}~\bibnamefont{F{\"o}lling}},
  \bibinfo{author}{\bibfnamefont{M.}~\bibnamefont{Feld}},
  \bibinfo{author}{\bibfnamefont{U.}~\bibnamefont{Schnorrberger}},
  \bibinfo{author}{\bibfnamefont{A.~M.} \bibnamefont{Rey}},
  \bibinfo{author}{\bibfnamefont{A.}~\bibnamefont{Polkovnikov}},
  \bibinfo{author}{\bibfnamefont{E.}~\bibnamefont{Demler}},
  \bibinfo{author}{\bibfnamefont{M.}~\bibnamefont{Lukin}}, \bibnamefont{and}
  \bibinfo{author}{\bibfnamefont{I.}~\bibnamefont{Bloch}},
  \bibinfo{journal}{Science} \textbf{\bibinfo{volume}{319}},
  \bibinfo{pages}{295} (\bibinfo{year}{2008}).

\bibitem[{\citenamefont{Blaizot and Ripka}(1986)}]{blaizot1986quantum}
\bibinfo{author}{\bibfnamefont{J.-P.} \bibnamefont{Blaizot}} \bibnamefont{and}
  \bibinfo{author}{\bibfnamefont{G.}~\bibnamefont{Ripka}},
  \emph{\bibinfo{title}{Quantum theory of finite systems}},
  vol.~\bibinfo{volume}{3} (\bibinfo{publisher}{MIT press Cambridge},
  \bibinfo{year}{1986}).

\bibitem[{\citenamefont{Puddy et~al.}(2015)\citenamefont{Puddy, Smith, Al-Taie,
  Chong, Farrer, Griffiths, Ritchie, Kelly, Pepper, and
  Smith}}]{puddy_multiplexed_2015}
\bibinfo{author}{\bibfnamefont{R.~K.} \bibnamefont{Puddy}},
  \bibinfo{author}{\bibfnamefont{L.~W.} \bibnamefont{Smith}},
  \bibinfo{author}{\bibfnamefont{H.}~\bibnamefont{Al-Taie}},
  \bibinfo{author}{\bibfnamefont{C.~H.} \bibnamefont{Chong}},
  \bibinfo{author}{\bibfnamefont{I.}~\bibnamefont{Farrer}},
  \bibinfo{author}{\bibfnamefont{J.~P.} \bibnamefont{Griffiths}},
  \bibinfo{author}{\bibfnamefont{D.~A.} \bibnamefont{Ritchie}},
  \bibinfo{author}{\bibfnamefont{M.~J.} \bibnamefont{Kelly}},
  \bibinfo{author}{\bibfnamefont{M.}~\bibnamefont{Pepper}}, \bibnamefont{and}
  \bibinfo{author}{\bibfnamefont{C.~G.} \bibnamefont{Smith}},
  \bibinfo{journal}{Applied Physics Letters} \textbf{\bibinfo{volume}{107}},
  \bibinfo{pages}{143501} (\bibinfo{year}{2015}), ISSN
  \bibinfo{issn}{0003-6951, 1077-3118}.

\bibitem[{\citenamefont{Veldhorst et~al.}(2016)\citenamefont{Veldhorst, Eenink,
  Yang, and Dzurak}}]{veldhorst2016silicon}
\bibinfo{author}{\bibfnamefont{M.}~\bibnamefont{Veldhorst}},
  \bibinfo{author}{\bibfnamefont{H.}~\bibnamefont{Eenink}},
  \bibinfo{author}{\bibfnamefont{C.}~\bibnamefont{Yang}}, \bibnamefont{and}
  \bibinfo{author}{\bibfnamefont{A.}~\bibnamefont{Dzurak}},
  \bibinfo{journal}{arXiv preprint arXiv:1609.09700}  (\bibinfo{year}{2016}).

\bibitem[{\citenamefont{Mead and Papanicolaou}(1984)}]{mead1984maximum}
\bibinfo{author}{\bibfnamefont{L.~R.} \bibnamefont{Mead}} \bibnamefont{and}
  \bibinfo{author}{\bibfnamefont{N.}~\bibnamefont{Papanicolaou}},
  \bibinfo{journal}{J. Math. Phys.} \textbf{\bibinfo{volume}{25}},
  \bibinfo{pages}{2404} (\bibinfo{year}{1984}).

\bibitem[{\citenamefont{Han et~al.}(2016)\citenamefont{Han, Malioutov, Avron,
  and Shin}}]{han2016approximating}
\bibinfo{author}{\bibfnamefont{I.}~\bibnamefont{Han}},
  \bibinfo{author}{\bibfnamefont{D.}~\bibnamefont{Malioutov}},
  \bibinfo{author}{\bibfnamefont{H.}~\bibnamefont{Avron}}, \bibnamefont{and}
  \bibinfo{author}{\bibfnamefont{J.}~\bibnamefont{Shin}},
  \bibinfo{journal}{arXiv:1606.00942}  (\bibinfo{year}{2016}).

\bibitem[{\citenamefont{Eisler and Zimbor{\'a}s}(2015)}]{eisler2015partial}
\bibinfo{author}{\bibfnamefont{V.}~\bibnamefont{Eisler}} \bibnamefont{and}
  \bibinfo{author}{\bibfnamefont{Z.}~\bibnamefont{Zimbor{\'a}s}},
  \bibinfo{journal}{New Journal of Physics} \textbf{\bibinfo{volume}{17}},
  \bibinfo{pages}{053048} (\bibinfo{year}{2015}).

\bibitem[{\citenamefont{Coser et~al.}(2016)\citenamefont{Coser, Tonni, and
  Calabrese}}]{coser2016towards}
\bibinfo{author}{\bibfnamefont{A.}~\bibnamefont{Coser}},
  \bibinfo{author}{\bibfnamefont{E.}~\bibnamefont{Tonni}}, \bibnamefont{and}
  \bibinfo{author}{\bibfnamefont{P.}~\bibnamefont{Calabrese}},
  \bibinfo{journal}{Journal of Statistical Mechanics: Theory and Experiment}
  \textbf{\bibinfo{volume}{2016}}, \bibinfo{pages}{033116}
  (\bibinfo{year}{2016}).

\bibitem[{\citenamefont{Aubrun}(2012)}]{aubrun2012partial}
\bibinfo{author}{\bibfnamefont{G.}~\bibnamefont{Aubrun}},
  \bibinfo{journal}{Random Matrices: Theory Appl.}
  \textbf{\bibinfo{volume}{1}}, \bibinfo{pages}{1250001}
  (\bibinfo{year}{2012}).

\bibitem[{\citenamefont{Fukuda and {\'S}niady}(2013)}]{fukuda2013partial}
\bibinfo{author}{\bibfnamefont{M.}~\bibnamefont{Fukuda}} \bibnamefont{and}
  \bibinfo{author}{\bibfnamefont{P.}~\bibnamefont{{\'S}niady}},
  \bibinfo{journal}{J. Math. Phys.} \textbf{\bibinfo{volume}{54}},
  \bibinfo{pages}{042202} (\bibinfo{year}{2013}).

\bibitem[{\citenamefont{Bergstra et~al.}(2013)\citenamefont{Bergstra, Yamins,
  and Cox}}]{bergstra2013making}
\bibinfo{author}{\bibfnamefont{J.}~\bibnamefont{Bergstra}},
  \bibinfo{author}{\bibfnamefont{D.}~\bibnamefont{Yamins}}, \bibnamefont{and}
  \bibinfo{author}{\bibfnamefont{D.}~\bibnamefont{Cox}}, in
  \emph{\bibinfo{booktitle}{International Conference on Machine Learning}}
  (\bibinfo{year}{2013}), pp. \bibinfo{pages}{115--123}.

\bibitem[{\citenamefont{Chollet et~al.}(2015)}]{chollet2015keras}
\bibinfo{author}{\bibfnamefont{F.}~\bibnamefont{Chollet}} \bibnamefont{et~al.},
  \emph{\bibinfo{title}{Keras}},
  \bibinfo{howpublished}{\url{https://github.com/fchollet/keras}}
  (\bibinfo{year}{2015}).

\end{thebibliography}

\clearpage

\onecolumngrid
\appendix

\section*{Appendix}

\renewcommand{\theequation}{S\arabic{equation}}
\renewcommand{\thefigure}{S\arabic{figure}}
\setcounter{equation}{0}
\setcounter{figure}{0}

\subsubsection{Measuring Moments in spin-$1/2$ Systems}

Here we show that the operational procedure described in the main text allows us to measure the moments $\mu_m$ for spin-1/2 systems.
Let $S^{c,d}_X$ be the operator that swaps copies $c$ and $d$ on subsystem $X$, which can be written as
$S^{c,d}_X = \prod_{j \in X} \Xi^{c,d}_j$ where
$
\Xi^{c,d}_{j} =
(\openone + \boldsymbol{\sigma}^{j,c} {\cdot} \boldsymbol{\sigma}^{j,d})/2$.
The projective measurement of $\Xi^{c,d}_j$ corresponds to a singlet triplet measurement (ST-measurement) between spins sitting at the same site $j$, but different copies $c$ and $d$.
Indeed, $\Xi^{c,d}_j$ has an outcome $-1$ for the singlet state and $1$ for the triplet states.
In view of this we write $S^{c,d}_X {=} \Pi^{c,d}_+ {-}\Pi_-^{c,d}$
where $\Pi^{c,d}_{\pm}$ correspond to the eigen-projections with corresponding eigenvalues $\pm 1$.

We now consider the case $m{=}3$ and then generalise to arbitrary $m$.
We first perform a sequential set of ST-measurements on copies $(1,2)$, with
outcome $\beta_{1}$ and then
do the same measurement on copies $(2,3)$, with outcome $\beta_2$.
We introduce the notation $S_X^{2,3} \circ S_X^{1,2}$ to describe this process.
After the first measurement, the (non-normalized) state of the system will be
$\Pi^{1,2}_{\beta_1} \rho^{\otimes 3} \Pi_{\beta_1}^{1,2}$,
while after the two sets of
measurements it is
$\Pi^{2,3}_{\beta_2} \Pi^{1,2}_{\beta_1} \rho^{\otimes 3} \Pi_{\beta_1}^{1,2}\Pi^{2,3}_{\beta_2} $.
Therefore,
\begin{align}
  \left\langle S_X^{2,3}\circ S_X^{1,2} \right\rangle &= \sum_{\beta_2}\sum_{\beta_1} \beta_1 \beta_2 \Tr\left[
  \Pi^{2,3}_{\beta_2} \Pi^{1,2}_{\beta_1} \rho^{\otimes 3} \Pi_{\beta_1}^{1,2}\Pi^{2,3}_{\beta_2}\right]
\cr&
  = \sum_{\beta_1} \beta_1 \Tr\left[\Pi_{\beta_1}^{1,2}
  S^{2,3}_{X} \Pi^{1,2}_{\beta_1} \rho^{\otimes 3} \right]
\cr&
  = \frac12\left(\Tr[S_{X}^{1,2}
  S^{2,3}_{X} \rho^{\otimes 3} ] +
   \Tr[S^{2,3}_{X} S^{1,2}_{X} \rho^{\otimes 3} ]\right)
\cr&
  = \Tr(\rho^{3}),
  \label{S1}
\end{align}
where we used the identity $\Pi_{\pm}^{c,d} {=}
(\openone{\pm}{S}_X^{c,d})/2$.

We now generalize the above argument for higher values of $m$.
We apply sequential ST-measurements on neighbouring copies, using the notation
$S^{m,m-1}_X{\circ} \cdots \circ S^{2,3}_X{\circ} S^{1,2}_X$, meaning that we first
perform $S^{1,2}_X$ and so forth.
Taking the averages one then finds that
$$\left\langle S^{m{-}1,m}_X{\circ}{\cdots}{\circ} S^{1,2}_X\right\rangle{=} \Tr[\mathcal
    P^{m{-}1,m}[\cdots\mathcal
        P^{23}[\mathcal P^{12}[\rho^{\otimes m}] ]]\cdots],
$$ where
$\mathcal P^{j,j{+}1}[\rho]{=} \sum_{\beta_j} \beta_j
\Pi^{j,j{+}1}_{\beta_j}\rho\Pi^{j,j{+}1}_{\beta_j}$.
We define the operators $S_X^{a,b,c,\dots}$ recursively as:
$\mathcal P^{j{+}1,j}[S_X^{j,a,b,\dots}]= [
S_X^{j{+}1,j,a,\dots}{+} S_X^{j,j{+}1,a,\dots}]/2 $.
Then, using the cyclic property of the trace, one finds that
$\left \langle
S^{m{-}1,m}_X{\circ}{\cdots}{\circ} S^{1,2}_A\right\rangle
{=}2^{2{-}m}
\sum_\kappa
\left\langle
S_X^{\kappa}
\right\rangle
$ where the $\kappa$ are $2^{m{-}2}$ different cyclic permutation of the elements $1,\dots,{m}$. For instance, for
$m{=}3$ one has $\kappa{=}\{123,132\}$.
In view of the above, this sequential set of ST-measurements corresponds to the measurement of the operator $\mathbb{P}^m_X=2^{2{-}m} \sum_\kappa S_X^{\kappa}$.

In summary, when $X$ is equal to both $A$ and $B$, one obtains the operator, $\mathbb{P}^m$ defined in Eq.~\eqref{S2}.
As proven above, such an operator is defined according to the following recursion relations~\cite{banchi2016entanglement}:
$
\mathbb{P}^m = \left(S_{A}^{m,m-1} S_{B}^{m,m-1} \mathbb{P}^{m-1} + \mathrm{h.c.} \right) / 2.
$
Moreover, the effect of the partial transpose on the recursion relation is as follows:
$
(\mathbb{P}^{m})^{T_B} = \left(S_{A}^{m,m-1} (\mathbb{P}^{m - 1})^{T_B} S_{B}^{m,m-1} + \mathrm{h.c.}\right)/2.
$
We take $m=3$ as an example:
\begin{align}
  (\mathbb{P}^{3})^{T_B}
  &=
  \left(S_{A}^{3,2} S_{A}^{2,1}\otimes S_{B}^{2,1} S_{B}^{3,2}
  + \mathrm{h.c.}\right)/2.
  \label{S2}
\end{align}
From which it can be seen, as described in the main text, that the order of measurements on $A$ and $B$ is reversed.
Indeed, for $A$, the ST-measurement is performed between copies 1 and 2, then 2 and 3, whereas for $B$, the ST-measurements is performed for copies 2 and 3, then 1 and 2.
Because of the non-commutative nature of these measurements, this ordering is crucial in order to yield moments of the partially transposed state as in Eq.~\eqref{S2}.

From the above derivation we find that
$\mu_m = \Tr\left[\rho_{AB}^{\otimes m} \mathbb{P}_m\right]$.
The variance of the above measurement can be found using the procedure of \eqref{S1},
noting that
$\Tr\left[\Pi_{\beta_1}^{1,2}
  S^{2,3}_{X} \Pi^{1,2}_{\beta_1} \rho^{\otimes 3} \right]$ is the provability of getting
  the outcome sequence $\beta_1$, $\beta_2$.
Therefore the variance is
\begin{align}
  (\Delta \mu_m)^2 &= \left(
  \sum_{\beta_m}\cdots\sum_{\beta_1} \beta_1^2\cdots \beta_m^2 \Tr\left[
    \Pi^{m-1,m}_{\beta_{m-1}} \cdots \Pi^{1,2}_{\beta_1} \rho^{\otimes m} \Pi_{\beta_1}^{1,2}\cdots
  \Pi^{m-1,m}_{\beta_{m-1}}\right]\right) - \mu_m^2
  = \cr &=
\left(
  \sum_{\beta_m}\cdots\sum_{\beta_1}  \Tr\left[
    \Pi^{m-1,m}_{\beta_{m-1}} \cdots \Pi^{1,2}_{\beta_1} \rho^{\otimes m} \Pi_{\beta_1}^{1,2}\cdots
\Pi^{m-1,m}_{\beta_{m-1}}\right]\right) - \mu_m^2 = \cr &= 1- \mu_m^2,
\end{align}
where we used the fact that $\beta_i^2=1$ and
$\sum_{\beta_i}   (\Pi^{i,i+1}_{\beta_{i}})^2 =
\sum_{\beta_i}  \Pi^{i,i+1}_{\beta_{i}} = \openone$. Repeating the experiment $R$ times we find
the standard deviation
\begin{equation}
  \Delta \mu_m = \sqrt{\frac{1-\mu_m^2}R}~.
  \label{stdev}
\end{equation}

\bigskip
\subsubsection{Measuring Moments in Bosonic Systems}

We show here how to measure the moments as given in Eq.~\eqref{eq:moments} for a bosonic system.
Unlike for the case of spin systems, we directly choose the operator $\mathbb{P}_m$ as a product of specific, non-hermitian, permutations, $\pi$, such that
\begin{equation} \label{eq:bosonperm}
\mathbb{P}_{m}^{T_B} = \bigotimes_{j \in A} V_{j,\pi} \bigotimes_{j\in B} V_{j, \pi}^T
\end{equation}
where
$V_{j,\pi} = \sum_{\{n_{j,c}\}} \ket{n_{j,1},\dots,n_{j,m}}\bra{\pi(n_{j,1}),\dots,\pi(n_{j,m})}$,
and $n_{j,c}=0,\dots,\infty$ labels the number of bosons in copy $c$ and physical site $j$.
We can write this operator in second quantized form as
\begin{equation}
    V_{j, \pi} = :e^{\sum_c a^{\dag}_{j, c} a_{j,\pi(c)} - a^{\dag}_{j,c} a_{j,c}}:
\end{equation}
where $:O:$ denotes the normal ordering of the operator $O$, and $a_{j,c}$ denotes the annihilation operator acting on site $j$ and copy $c$.
We choose $\pi$ as the shift permutation such that $\pi(c) = c+1$.
Note that $V_{j,\pi}^T=V_{j,\pi^{-1}}$.
In order to diagonalise $V_{j,\pi}$ we introduce the Fourier transform, which acts independently on each site $j$ as
\begin{align}\label{eq:fourier}
    \tilde{a}_{j,c} &= \frac{1}{\sqrt{m}} \sum_{c'=0}^{m-1} e^{+\frac{i 2 \pi}{m} c c'} a_{j,c'}, & \text{for } &j \in A,\cr
    \tilde{a}_{j,c} &= \frac{1}{\sqrt{m}} \sum_{c'=0}^{m-1} e^{-\frac{i 2 \pi}{m} c c'} a_{j,c'}, & \text{for } &j \in B.
\end{align}
After such a transformation, both the operators $V_{j,\pi}$ for $j\in A$ and $V_{j,\pi^{-1}}$ for $j \in B$, take the form
$
{:}e^{\sum_c (e^{\frac{i 2 \pi}{m} c} - 1) \tilde{a}^{\dag}_{j,c} \tilde{a}_{j,c}}{:}.
$
The normal ordering can be removed by using the identity \cite{blaizot1986quantum}
 ${:}e^{(e^{\lambda} - 1)a^\dag a}{:} = e^{\lambda a^\dag a}$ bringing Eq.~\eqref{eq:bosonperm} to the form:
\begin{equation}
\mathbb{P}_{m}^{T_B} = \prod_{j\in\{A,B\},c} e^{\frac{i 2 \pi c}{m} \tilde{a}^{\dag}_{j,c} \tilde{a}_{j,c}}.
\end{equation}
The expectation value in Eq.\eqref{eq:moments} can thus be measured in three steps.
First, perform the Fourier (inverse Fourier) transform between copies at the sites belonging to $A$ ($B$), as written in Eq.~\eqref{eq:fourier}.
Second, measure the bosonic occupation number with outcome $n_{j,c}$ at every site and compute the outcome of the permutation operator as $\phi =e^{\sum_{j\in\{A,B\},c}\frac{i 2 \pi c}{m} n_{j,c}}$.
Finally, compute the expectation value as the average over many repeatations  of the above steps:
$
\mu_m = \langle \phi \rangle
$.

The standard deviation can obtained from the fact that $\phi^* = \phi$. As such
$(\Delta \mu_m)^2 = \langle{\phi\phi^*}\rangle-\langle{\phi}\rangle^2 = 1-\mu_m^2$ and
after $R$ repetitions we find the same result of Eq.~\eqref{stdev}.

\bigskip
\subsubsection{Experimental Feasibility.}
Our proposal for measuring logarithmic negativity can be realised in different physical set-ups.
In solid state quantum technologies the possibility of obtaining multiple copies is rapidly emerging.
For example, in~\cite{puddy_multiplexed_2015} a system of two parallel quantum dot arrays, each with 7 sites, was realised in GaAs.
Alternatively, in silicon quantum technologies, recent proposals have put forward schemes for large two-dimensional arrays of quantum dots exploiting the well-established CMOS technology~\cite{veldhorst2016silicon}.
Preparing multiple copies of strongly correlated many-body systems in such structures would be a natural capability.
In these quantum dot arrays, singlet-triplet measurements are already well developed~\cite{petta2005coherent}, using either charge detection through quantum point contacts, or capacitance measurement through radio-frequency reflectometry.
Since the singlet and triplet states correspond to the anti-symmetric and symmetric subspaces respectively, they are the eigenstates of the swap operator and thus the singlet-triplet measurement outcome can be mapped to the swap measurement outcome
.
The full details of the above, and its generalization to multiple consecutive swaps, are available elsewhere in the appendix.

In optical lattices the situation is currently even more advanced, and
multiple copies of many-body systems can be easily isolated from each other in adjacent lattice rows.
However, the counterpart to the solid state swap measurement in such systems is destructive.
This means that consecutive swaps cannot be measured, limiting the number of copies to only two.
Nonetheless, as explained above, our procedure is valid for whenever forward-backward measurements that are equivalent to permutations are available.
In the case of optical lattices, the Fourier transform provides such a capability that has also been experimentally implemented~\cite{islam2015measuring}.
This is performed by tilting alternating rows of the lattice such that the bosons undergo a series of effective beam-splitters~\cite{daley2012measuring}.
In our case, this operation would need to be performed on two subsystems in forwards and backwards fashion -- the formal equivalence of this to two permutations is also detailed in the appendix.
Combining these above experimental techniques for measuring the partially transposed moments, with our machine learning method for extracting the logarithmic negativity, provides a realistic and complete scheme for accurately estimating entanglement in general many-body systems.

One potential source of error experimentally is that the copies may not be perfectly identical.
However, it can be shown that small deviations in the fidelity of copies leads to only small changes in the moments.
Since both the Chebyshev and neural network approaches yield entanglement estimators that are smooth functions of the moments, the overall error is therefore well controlled.

\bigskip
\subsubsection{Chebyshev Expansion}

\begin{figure}[tb]
    \centering
    \includegraphics[width=0.6\linewidth]{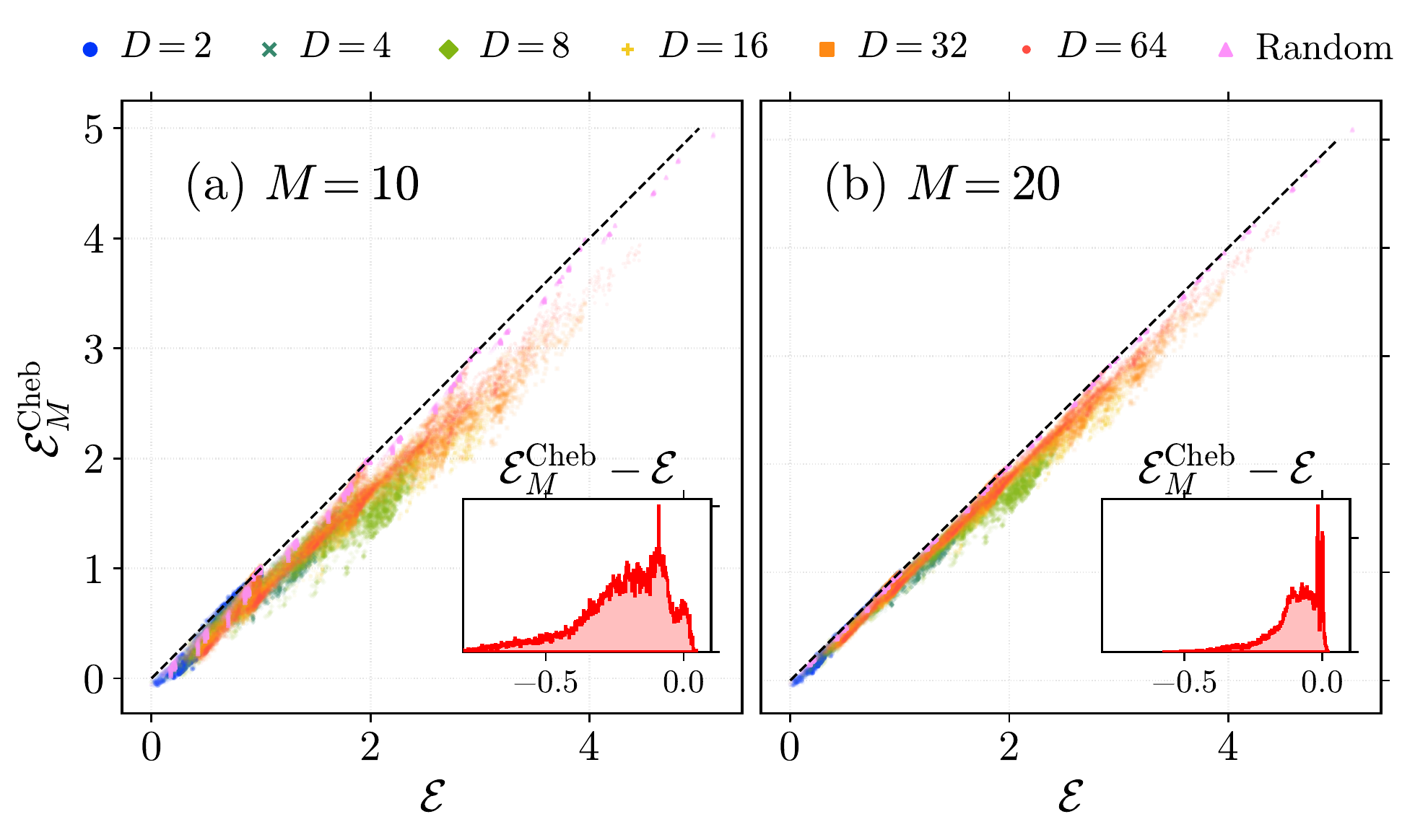}
    \caption{\label{fig:cheb-logneg-approx}
    \textbf{Chebyshev approximation for estimating entanglement.}
    Estimated logarithmic negativity $\mathcal{E}^{\mathrm{Cheb}}_{M}$, using the Chebyshev approximation vs. actual logarithmic negativity $\mathcal{E}$, for a wide range of random states and partitions.
    We use both R-GPS and R-MPS states with varying bond-dimension $D$, and various system sizes $N_A$, $N_B$ and $N_C$, such that $N_A + N_B \le 12$ and the total length $N \le 24$.
    The Chebyshev approximation is calculated using the moments $\mu_m$ generated from: (a) $M=10$ copies; (b) $M=20$ copies.
    The respective insets show the distribution of error, $\mathcal{E}^{\mathrm{Cheb}}_M - \mathcal{E}$.
    }
\end{figure}

In this section, we demonstrate an analytical method, based on functional approximation, for estimating the logarithmic negativity from the information contained in the moments, $\mu_m$.

Since logarithmic negativity, $\mathcal{E}$, is a function of the eigenvalues $\{\lambda_k\}$, one could try to directly reconstruct the main features of the spectrum $\{\lambda_k\}$ using a few measured moments $\{\mu_m\}$ -- an approach closely related to general Hausdorff moments problem in statistics~\cite{mead1984maximum}.
Nonetheless, it is known from a numerical perspective that the Hausdorff problem is unstable~\cite{viano1991solution}.
To avoid such instabilities, we might try an alternative approach based on functional approximation.
Considering that $\mathcal{E}=\log_2 \Tr f(\rho_{AB}^{T_B})$ with $f(x)=|x|$,
if we can find a polynomial expansion $f(x) \approx \sum_{m=0}^{M} \alpha_m x^m$, then by linearity of the trace, $\mathcal{E} = \log_2 \sum_{m=0}^{M} \alpha_m \mu_m$, with $\mu_m$ as given in Eq.~\eqref{eq:moments}.
In other words, given a polynomial expansion of the absolute function, $f(x)$, -- i.e. the coefficients $\alpha_m$ -- one can approximate the entanglement using a finite number of moments.
A naive choice for this would be a Taylor expansion, but the non-analyticity of $f(x)$ at $x=0$ prevents convergence.
On the other hand, a nearly optimal choice for approximating a function throughout an interval rather than around a point, is a Chebyshev expansion~\cite{han2016approximating}.
On the interval $[-1, 1]$, this yields $f(x) \approx \sum_{m=0}^{M} t_m T_m(x)$ where the $M + 1$ Chebyshev polynomials $T_{m}(x)$ are known $m$-th order polynomials.
The coefficients $t_m$ are given, via the orthogonality of $T_m(x)$, as
$t_m = \frac{2 {-} \delta_{m0}}{n {+} 1} \sum_{j=0}^{M} f(x_j)T_m(x_j)$, where $x_j=\cos\left[\pi(j {+} 1/2) / (M {+} 1)\right]$ are the Chebyshev nodes.
When $f$ is defined on a different interval, one can simply linearly transform the Chebyshev points and polynomials.
Although in principle the spectrum of a generic state $\rho_{AB}^{T_B}$ lies between $-1/2$ and $1$, in practice it is often much more tightly clustered.
Decreasing the window size significantly improves the approximation for a fixed $M$. Therefore, we need to find the minimal sized window such that all of the spectrum of a given $\rho_{AB}^{T_B}$ is contained.
A tight guess for such a window can be found since $\mu_m = \sum_k\lambda^m \ge \lambda_{\mathrm{max}}^{m}$ when $m$ is even, with $\lambda_{\mathrm{max}}$ being the eigenvalue with largest absolute value.
Thus in our numerics, we define the window as $[-a, a]$, with $a=\mu_{M}^{1/M}$
.
The quality of the Chebyshev approximation rapidly increases with $M$ and becomes exact in the limit $M\to\infty$.

In Fig.~\ref{fig:cheb-logneg-approx}(a) we plot the relationship between the real logarithmic negativity, $\mathcal{E}$, and the approximated value, $\mathcal{E}^{\mathrm{Cheb}}_M$, calculated using the Chebyshev expansion for $M=10$ copies.
The random states tested are the same as those considered in the main numerical results text.
As the figure shows, $\mathcal{E}^{\mathrm{Cheb}}_M$ typically underestimates $\mathcal{E}$, especially for MPS states.
As we discuss in the next section, we attribute this to the fact that the distribution of $\lambda_k$ is peaked around zero, particularly for MPS, where the Chebyshev approximation error is always negative.
In the inset we plot a histogram of the errors $\mathcal{E}^{\mathrm{Cheb}}_M - \mathcal{E}$ which clearly shows this negative bias.
As shown in Fig.~\ref{fig:cheb-logneg-approx}(b) and its inset, by doubling the number of copies to $M=20$, the accuracy of this method is significantly improved.
While $M=20$ is a significant experimental requirement, we emphasize that the Chebyshev approach provides an analytical tool for estimating $\mathcal{E}$ from a known expression of the moments.
In particular, when the moments can be expressed in compact form, such as for free fermionic systems~\cite{eisler2015partial,coser2016towards}, then the Chebyshev expansion can provide an analytic formula for $\mathcal{E}$.

\bigskip

\subsubsection{Generation of random states}

Random generic pure states (R-GPS) have been obtained by generating random vectors with complex elements distributed according to the normal distribution.
R-GPS obtained by sampling from the Haar measure have also been considered, though they are numerically more demanding. Nonetheless, they provide the same results.
Random Matrix Product States, R-MPS, have been obtained by writing
$\ket{\psi}=\sum_{\{i_j\}} \Tr[A^{(1), i_1} A^{(2), i_2} \dots A^{(N), i_N}] \ket{i_1,i_2,\dots,i_N}$ for random tensors $A^{(j), i}_{kl}$, where $j=1,\dots,N$, $i=0,1$, and $k,l=1,\dots,D$, being $D$ the bond dimension, with complex elements drawn from a normal distribution.

\bigskip


\begin{figure}[t]
    \centering
    \includegraphics[width=0.6\linewidth]{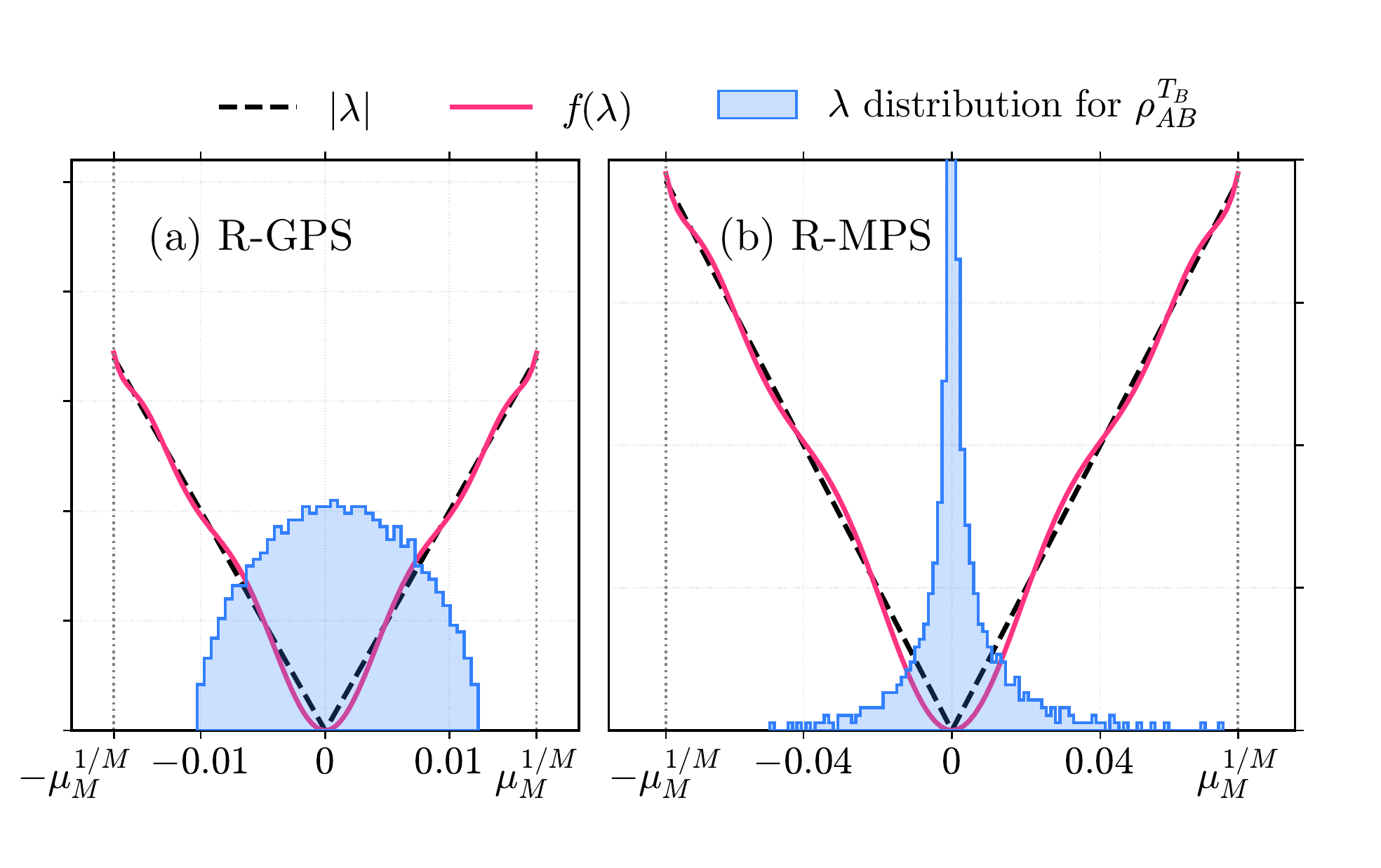}
    \caption{
    \textbf{Accuracy of the Chebyshev approximation.}
    Distribution of the eigenvalues $\{\lambda_k\}$ of $\rho_{AB}^{T_B}$ for (a) R-GPS and (b) R-MPS with $D=32$ with $N_A=5$, $N_B=5$ and $N_C=5$.
    The dashed lines show $|\lambda|$ while the solid lines show its Chebyshev approximation for $M=10$.
    Vertical dotted lines denote the estimated bounds on the spectrum calculated as $\pm \mu_M^{1/M}$.
    }
    \label{fig:spectrum-cheb-abs}
\end{figure}

\subsubsection{Chebyshev Approximation Error Analysis}

In this section we study in detail where the error in the Chebyshev approximation stems from and how it is related to the spectral properties of $\rho_{AB}^{T_B}$.
It is convenient to consider the thermodynamic limit ($N_A, N_B \gg 1$), where one can describe the spectrum of $\rho_{AB}^{T_B}$ as a probability density function, $\omega(\lambda)$, and write the logarithmic negativity of Eq.~\eqref{eq:negativity} as
\begin{equation} \label{eq:contin-neg}
    \mathcal{E} = \log_2 \int |\lambda| \omega(\lambda) d \lambda.
\end{equation}
Therefore, this underlying function $\omega(\lambda)$ determines $\mathcal{E}$, though it is not directly accessible from measurements of the moments $\mu_m$.
Nonetheless, from a theoretical perspective, the study of $\omega(\lambda)$ for certain classes of states yields insight into the performance of the Chebyshev approximation and machine learning approaches.

For R-GPS, $\ket{\Psi_{ABC}}$, as proven in \cite{aubrun2012partial,fukuda2013partial}, the spectral distribution $\omega(\lambda)$ tends towards a shifted Wigner semi-circle law in the limit of large $N_A$, $N_B$ and $N_C$.
This is described by the continuous distribution
$
\omega_{\mathrm{SC}}(\lambda) = \frac{d^2}{2 \pi \sigma^2} \sqrt{4\sigma ^2 - (d \lambda - 1)^2}
$
where $d = 2^{N_A {+} N_B}$ and $\sigma^2 = 2^{N_A {+} N_B {-} N_C}$, and $|d \lambda - 1| < 2 \sigma$.
An instance of the spectral distribution $\omega(\lambda)$ is shown in Fig.~\ref{fig:spectrum-cheb-abs}(a) for finite values of $N_A$, $N_B$ and $N_C$, where one can already see a clear semi-circular shape.
As is also evident from the figure, the support of the distribution is far smaller than the theoretical interval $\lambda_k \in [-1/2, 1]$, yet the bound established above as $|\lambda_k| \le \mu_M^{1/M}$ quite tightly captures the real interval.

On the other hand, random states constructed using a MPS ansatz~\cite{schollwock2011density} with fixed bond dimension $D \ll 2^{N_A + N_B + N_C}$ -- which inherently obey an area-law -- show a significantly different distribution $\omega(\lambda)$, with a high concentration around 0 but long tails on either side.
This can be seen in Fig.~\ref{fig:spectrum-cheb-abs}(b) for a single R-MPS instance of $\psi_{ABC}$.
Nonetheless, the support of this type of distribution is even more tightly bounded by $|\lambda_k| \le \mu_M^{1/M}$.

In this thermodynamic limit the Chebyshev approximation can be understood from Eq.~\eqref{eq:contin-neg}, as $|\lambda| \to f(\lambda)$.
The support of $\omega(\lambda)$ for the two classes of states discussed above is typically very different, being much wider for R-MPS.
The effect of this wider range can be seen if we consider the error in the Chebyshev approximation, $f(\lambda) - |\lambda|$, as a function of $\lambda$.
By construction this error is spread roughly throughout the interval, with alternating sign.
In the case of $\omega(\lambda)$ for R-MPS, the peak at zero, together with the large support, concentrate a large number of eigenvalues into a small region with the same signed error.
This gives a negative bias that does not exist when $\omega(\lambda)$ is more even throughout the interval, as for random pure states.
Therefore, Chebyshev methods are expected to have a larger error for area law states.

\bigskip
\subsubsection{Neural Network Details}

In deep neural networks, the unknown function $f$ mapping inputs to outputs is approximated with a directed graph organized in layers, where the first layer is the input data and the last one is the output.
In our case, the input data consists of the numbers $(N_A, N_B, \mu_2, \dots, \mu_M)$, namely, the number of spins in each subsystem and the non-trivial moments.
The value $s_k^{(\ell)}$ of the
$k$-th node in layer $\ell$ is updated via the equation
$  s^{(\ell)}_k {=} A^{\ell}\left[\sum_j
w_{kj}^{(\ell{-}1)}
s^{(\ell{-}1)}_j\right]~,$
where $A^{\ell}$ is an appropriate (typically non-linear) activation function
and $w_{kj}^{(\ell{-}1)}$ is the weight between node $k$ in layer $\ell$
and node $j$ in $\ell{-}1$.
The training procedure consists in finding the optimal weights
$w$ by minimizing a suitable {\it cost function}.

In our numerical investigations, we use the Hyperopt~\cite{bergstra2013making} and Keras~\cite{chollet2015keras} packages to find the optimal network structure, including the number of hidden layers.
For example, for $M=3$, the resulting network consists of two hidden layers, both with rectified linear unit (ReLU) activation functions, with $100$ and $56$ neurons respectively.
For $M=10$ however, the resulting network consists of three hidden layers, with exponential linear unit (ELU), ReLU and linear activation functions, with $61$, $87$ and $47$ neurons respectively.

\bigskip

\subsubsection{Neural Network Error Analysis}
We have seen that neural networks can provide very accurate predictions for the logarithmic negativity given only a few moments.
However, in a experimental setting it is also important to be aware of the effective size of any error bars.
We note first of all, that the question of neural networks themselves providing error estimates is an important yet open problem.
On the other hand, broad values for the error can be inferred statistically from the training process.
For example, the standard deviation in the error of the neural networks predictions for unseen test data is a good guide --- and this is exactly the standard deviation of the distributions plotted in the insets of Fig.~\ref{fig:ml-logneg-approx}.
One could also calculate the standard deviation as a function of $\mathcal{E}$ (i.e. windowed between $\mathcal{E}{-}\delta, \mathcal{E}{+}\delta$, for a small value $\delta$) to give a more dynamic estimate.
As this data has never been used to optimise the neural network, there is every reason to associate similar error bounds for any new experimental data use to predict $\mathcal{E}$, as long it is qualitatively similar.
This similarity, in the example of the quantum quench shown in Fig.~\ref{fig:evo}, is achieved by training with both area- and volume-law entangled random states, yielding errors that roughly match those of the test data.

Another area of concern might be the effect of supplying slightly inaccurate moments, $\{\mu_m\}$, (due to imperfect copies or measurement error for example) to the neural network.
We note however that since the neural networks produce smooth functions of their variables, any errors are well controlled -- small errors will only produce small changes in the estimate.

\subsubsection{Neural Network Sensitivity}

Although the neural network produces a smooth output, there is the question of how much the predicted logarithmic negativity changes if the measured moments are imprecise.
We have shown very few moments are required to estimate the entanglement accurately, and these can be measured with an effort that scales linearly with the subsystem sizes.
However, to be scalable, it is necessary that the number of repeat measurements, $R$, which sets the standard error $\Delta \mu_m$ via \eqref{stdev}, also scales efficiently, i.e., sub-exponentially.

\begin{figure}[tb]
    \centering
    \includegraphics[width=0.5\linewidth]{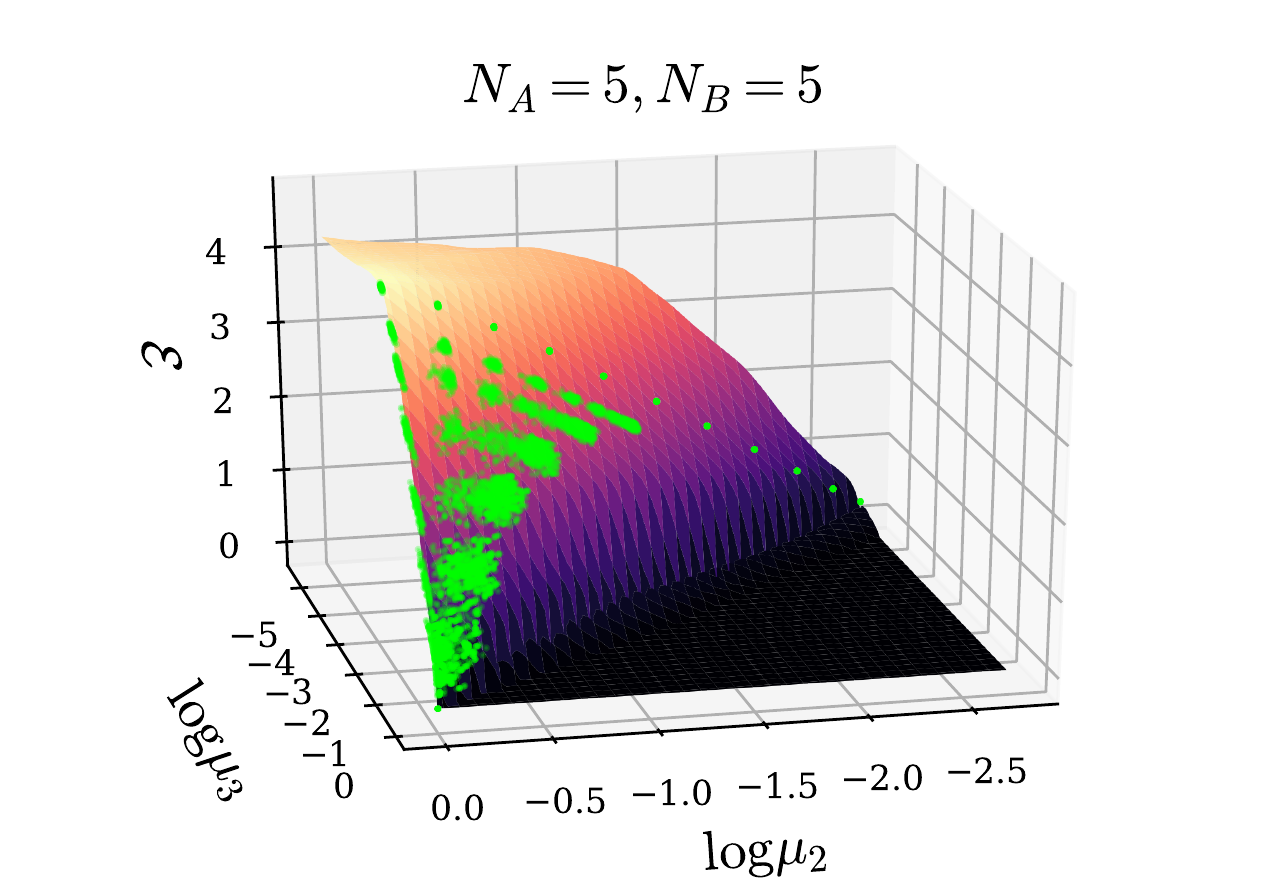}
    \caption{Entanglement as a function of $\mu_2$ and $\mu_3$ for the random set of density matrices with $N_A=N_B=5$. The green points represent the real logarithmic negativities, while the surface shows the neural network's predictions for the whole space.}
    \label{fig:neural-3d}
\end{figure}

To give a feeling for the dependence of the logarithmic negativity estimator on the the moments $\mu_m$, in Fig.~\ref{fig:neural-3d} we show a plot of $\mathcal{E}^\mathrm{ML}_\mathrm{M=3}$ as a function of $\mu_2$ and $\mu_3$ (for fixed $N_A$ and $N_B$ which are also inputs to the neural network).
In this plot, the scatter points are real data from the random training set, while the surface is effective function of the neural network.
As can be seen, the neural network is smooth (apart from the enforced lower bound of 0), with the vast majority of data overlaying areas with reasonable gradient.

\begin{figure}[tb]
    \centering
    \includegraphics[width=0.4\linewidth]{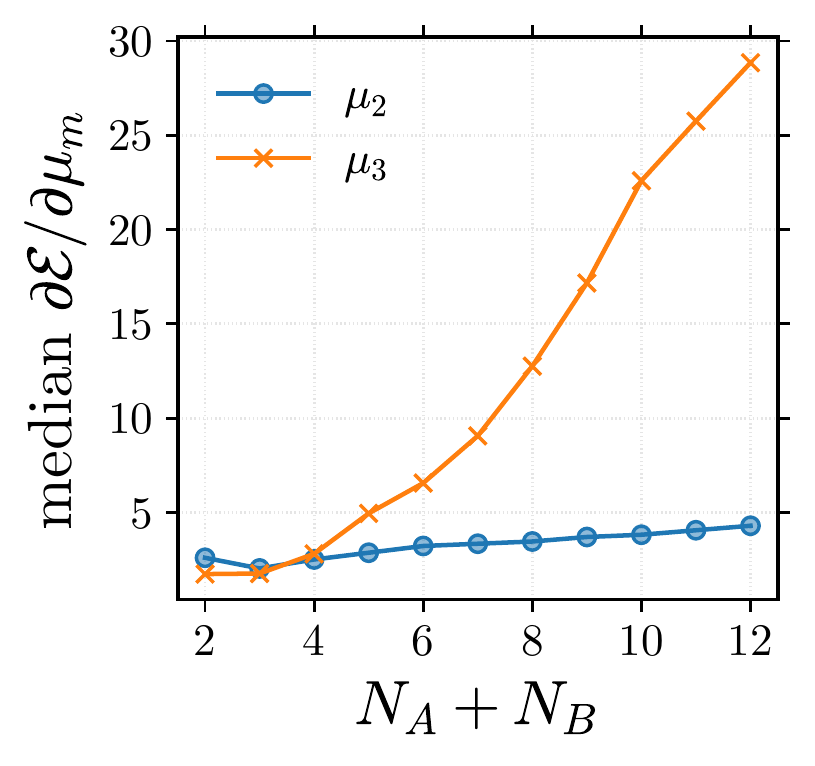}
    \caption{Typical gradient of the neural network estimator with respect to the measured moments $\mu_2$ and $\mu_3$.}
    \label{fig:nn-median-gradient}
\end{figure}

The error in the entanglement estimation is given by the gradient of the neural network estimator with respect to the set of input moments $\{\mu_m\}$, multiplied by the measurement error, namely:
\begin{equation}
    \Delta \mathcal{E} \approx \sum_m \Delta\mu_m \partial \mathcal{E} / \partial \mu_m,
\end{equation}
where the partial derivative denotes the neural network sensitivity.
Since $\Delta\mu_m \propto R^{-1/2}$, as $N_{A+B}$ increases, it is required that the ratio of the sensitivity to $R$ remains manageable.
Specifically, this requires that the sensitivity does not diverge as the subsystem sizes grow.
To probe this, we take the set of random data and compute the gradient of $\mathcal{E}^{\mathrm{ML}}_{\mathrm{M=3}}$ with respect to $mu_2$ and $mu_3$ for each data point in the set.
In figure~\ref{fig:nn-median-gradient} we plot the typical gradient, the median, as a function of total system size $N_{A+B}$.
As is shown, there is a roughly linear increase in sensitivity with respect to both moments, with the absolute value in the manageable range $<30$, and more accuracy required for $\mu_3$.
This confirms that the number of required measurements, $R$ does not scale exponentially with the system sizes.

\subsection{Comparison with approximate state reconstruction methods}

Unlike our method, approximate polynomial state reconstruction schemes, e.g. based
tensor networks~\cite{cramer2010efficient,lanyon2016efficient} or neural
network states~\cite{torlai2018neural,torlai2018latent}, are normally focused on
finding an approximate representation of a state that accurately reproduces
experimental observables. Let $\rho_r$ be the experimentally reconstructed state
from a polynomial number of measurements and suppose that the expectation values
predicted by this state are ``close enough'' to those predicted by the true state
$\rho$. This means that
$
	\Tr[\rho_r A] \approx \Tr[\rho A],
	$
for any observable $A$. Alternatively we may ask that
$
	\Tr[\rho_r M_i] \approx \Tr[\rho M_i],
	$
	for any positive operator valued measurement (POVM) $M_i$. Because of Helstrom's theorem
\cite{bengtsson2006geometry} the maximal distance between the distributions obtained
from the true and reconstructed state is given by the trace distance
\begin{equation}
	D(\rho_r,\rho) = \frac12 \|\rho-\rho_r\|_1 = \frac12 \max_{\{M\}} \sum_i \Tr[M_i(\rho-\rho_r)]~.
\end{equation}
Therefore, if the states $\rho$ and $\rho_r$ are close with respect to the trace distance,
any expectation value obtained from the reconstructed state is close to the true value.
We may therefore assume that a ``good reconstruction'' has $D(\rho,\rho_r)\approx 0$.
Although this is true for any expectation value, entanglement measures are
non-analytic functions of expectation values and, as such, may display a higher
sensitivity to small errors in the reconstruction. Indeed, many entanglement measures
are not continuous, so
\begin{align}
	D(\rho,\sigma) &\to 0, &{\rm does~not~imply}&  &|\mathcal E(\rho)-\mathcal E(\sigma) |\to 0.
\end{align}
In fact, the entanglement negativity does not even satisfy the requirement of
asymptotic continuity (see table 15.2 in \cite{bengtsson2006geometry}). Consider
two $N$ qubit states $\rho$ and $\sigma$ such that $\|\rho-\sigma\|\to 0$ for
large $N$ (many particles). A measure if asymptotically continuous if
$\|\rho-\sigma\|_1\to 0$ implies $|\mathcal E(\rho)-\mathcal E(\sigma)|/\log(d_N) \to 0$
where $d_N=2^N$ is the Hilbert space dimension. Since the negativity does not satisfy
asymptotic continuity the difference between the entanglement of two `close' many-body state
can diverge faster than $\mathcal O(\log(d_N))\simeq \mathcal O(N)$ for large $N$.
Because of
this important point, the negativity obtained from states reconstructed with a polynomial
number of measurements can not be considered a valid approximation of the true negativity.

In contrast, our method avoids the intermediate step of reconstructing the quantum state
and uses a polynomial number of measurements for the sole purpose of reconstructing the
negativity. As shown in the previous section, this has the advantage that the prediction
is accurate and the error is well controlled.

\bigskip
\subsection{Additional Numerical Results}

In this section we further demonstrate the ability of the neural network, trained only with random states, to accurately predict the logarithmic negativity in several different classes of physical states.
These include states likely to be the most challenging for the neural network due to their high degree of symmetry, which is not enforced in the random training set.
We note that it is highly likely that these results could be improved even further with the adoption of specialized training sets, chosen to match the model under study. However, for the sake of generality, we focus here on training only with a set of random states with no underlying physical assumptions.

\subsubsection{Ground-states through a quantum phase transition}

Quantum phase transitions are a topic of interest for various fields of physics, noted for their display of various entanglement structures. Here we study the XX-model with transverse field:
\begin{equation}
    H_{\text{XX}} = \sum_{i=1}^{L - 1} \left( \sigma^X_i \sigma^X_{i + 1} + \sigma^Y_i \sigma^Y_{i + 1} \right) + B_Z \sum_{i=1}^{L} \sigma^Z_i ,
\end{equation}
where $B_Z$ denotes the magnetic field. This system undergoes a quantum phase transition at $B_Z = 1$, above which the system enters a phase with a separable ground-state.

\begin{figure}[tb]
    \centering
    \includegraphics[width=0.6\linewidth]{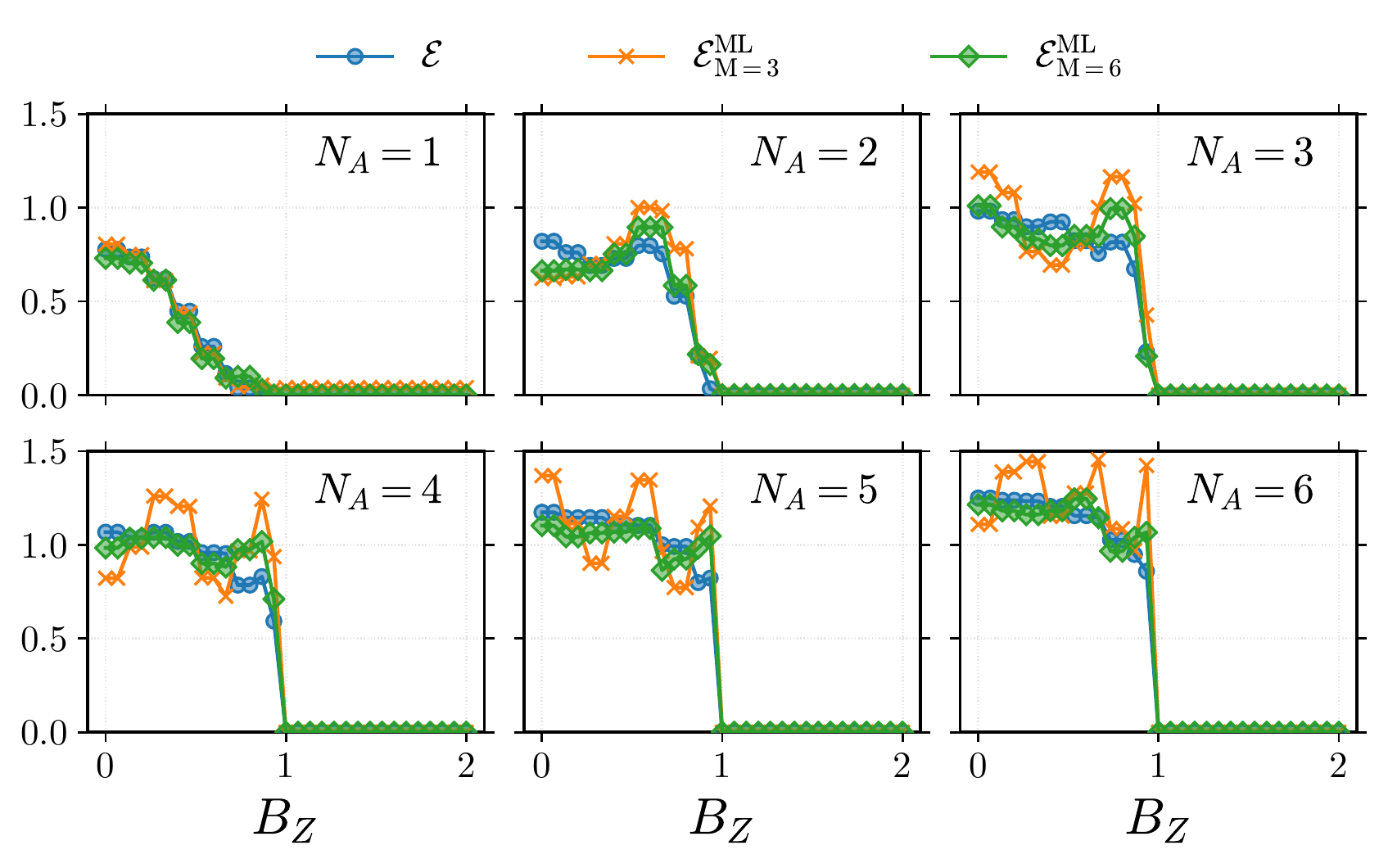}
    \caption{Entanglement estimation in the ground-state of the XX-model phase across the phase transition driven by transverse field, $B_Z$. The groundstate, of total length $L=20$ is tri-partitioned, with two adjacent subsystems of size $N_A=N_B$ and environment size $N_C=L - N_A + N_B$.
    The entanglement between $A$ and $B$ is then computed and estimated from $\rho_{AB}$.
    The blue, orange and green lines show the true logarithmic negativity, the neural network estimated quantity with 3 copies, and the neural network estimated quantity with 6 copies respectively.
    }
    \label{fig:phase-transition}
\end{figure}

In Fig.~\ref{fig:phase-transition} we show the logarithmic negativity, exact and predicted from moments using the neural network, for a variety of sizes of two adjacent blocks of spins within the groundstate of $H_{\text{XX}}$ across the transition.
The spin chain is divided into three blocks, of size $N_A$, $N_B$ and $N_C$.
System $C$ is traced out, then the entanglement is found between $A$ and $B$. This state is generally mixed as long as $N_C > 0$.
The neural network is exactly as defined in the main text --- trained solely with random states --- and we show results for using 3 and 6 moments (copies).
As can be seen from the figure, the key features of the transition are well reproduced in both cases, with the critical point ($B_Z=1$) clearly defined.
While there are some fluctuations in the quantitative estimate of $\mathcal{E}$
for 3 copies, these are significantly suppressed by raising the number of moments used to 6 copies.
It is remarkable that the neural network can capture the entanglement properties of these highly symmetric ground-states despite having only been trained with random states.

\subsubsection{W-state}

\begin{figure}[tb]
    \centering
    \includegraphics[width=0.6\linewidth]{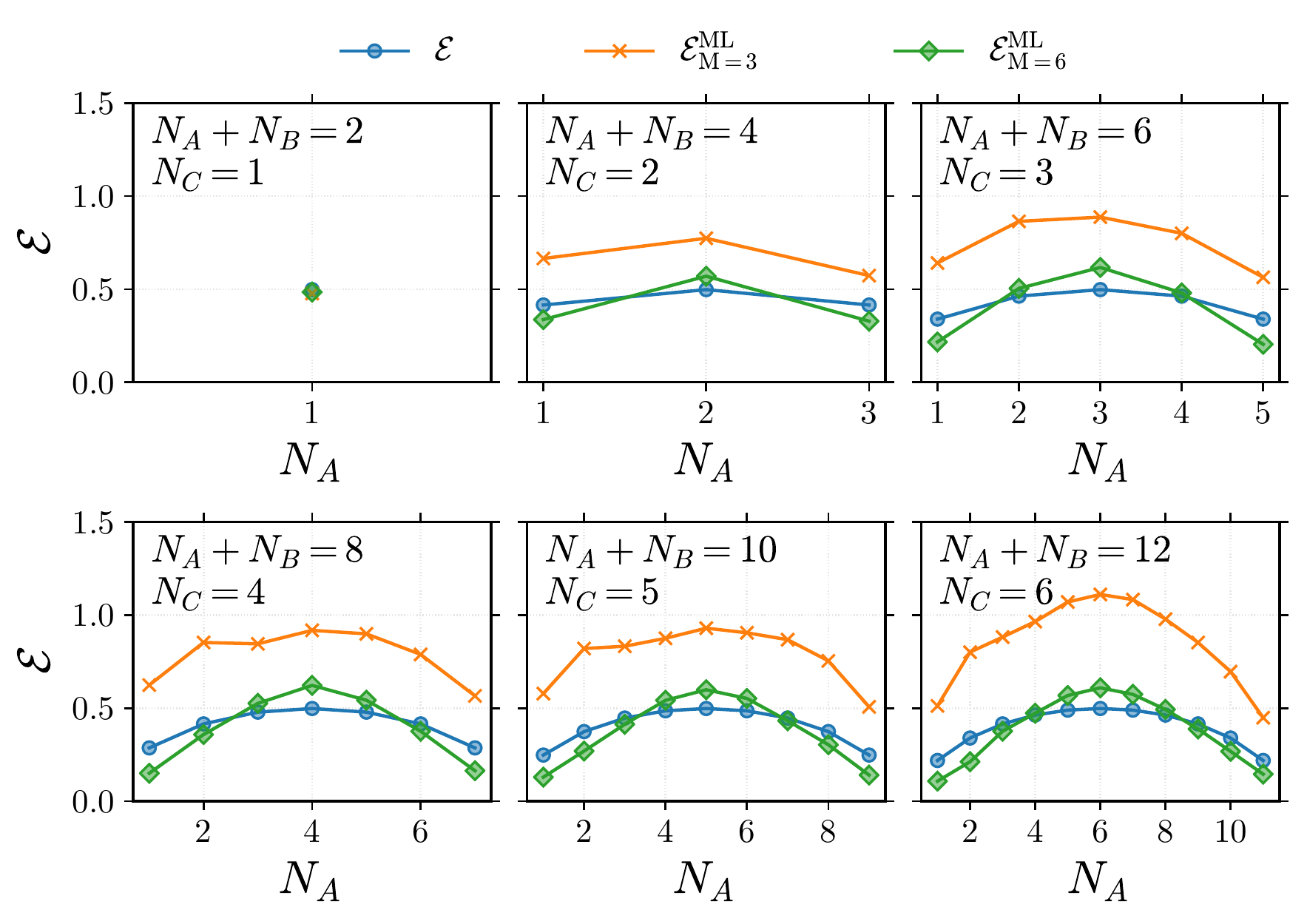}
    \caption{Entanglement estimation in the W-state. We show here a representative sample of the entanglement estimation in the W-state for various lengths ($L=N_{A+B} + N_C$) and partitions, as a function of subsystem $A$ size, $N_A$.
    The blue, orange and green lines show the true logarithmic negativity, the neural network estimated quantity with 3 copies, and the neural network estimated quantity with 6 copies respectively.}
    \label{fig:w-state}
\end{figure}

Our training set is composed of random states with no imposed symmetry, in which highly entangled states represent the largest class.
As such, low-entangled, highly symmetric states should generally pose the greatest challenge.
Here we study the paradigmatic example of the W-state, which is defined as:
\begin{equation}
    |W_L\rangle = \frac{1}{\sqrt{L}} \left (
    |100 \cdots 0 \rangle +
    |010 \cdots 0 \rangle + \cdots +
    |000 \cdots 1 \rangle \right)
\end{equation}
for $L$ qubits.
This state has a MPS representation with bond-dimension 2, making it one of the lowest possible entangled many-body states.
Among other symmetries, it is also fully permutationally symmetric, making it highly distinct from the random training set.
As in the previous sub-section, the W-state is divided into three blocks, of size $N_A$, $N_B$ and $N_C$.
System $C$ is traced out, then the entanglement is found between $A$ and $B$. This state is generally mixed as long as $N_C > 0$.
The choice of partitions is irrelevant due to the permutation symmetry.
In Fig.~\ref{fig:w-state} we plot the real logarithmic negativity and the neural network predictions using 3 and 6 copies respectively, trained as before solely with random states. We show a few representative combinations of $N_A$, $N_B$ and $N_C$.
Although the overall trend is well captured by both the 3-copy and 6-copy neural network, the absolute accuracy is only reasonable for the 6-copy scheme.
Given the atypicality of the W-state with respect to the training set, it is expected that this requirement for extra resources could be alleviated by training with more specialized states such as those with high degree of symmetry.
We also note our protocol works best for highly entangled state ($>1$ ebit of entanglement unlike the W-state here).
Nonetheless neural network is easily capable of identifying when the entanglement is relatively low ($<$ 1 ebit of entanglement).
In these cases, instead of resorting to more copies, it would also be feasible to switch to MPS-tomography~\cite{lanyon2016efficient}, which is efficient for low levels of entanglement.

\subsubsection{Quench across a phase transition}

\begin{figure}[tb]
    \centering
    \includegraphics[width=0.6\linewidth]{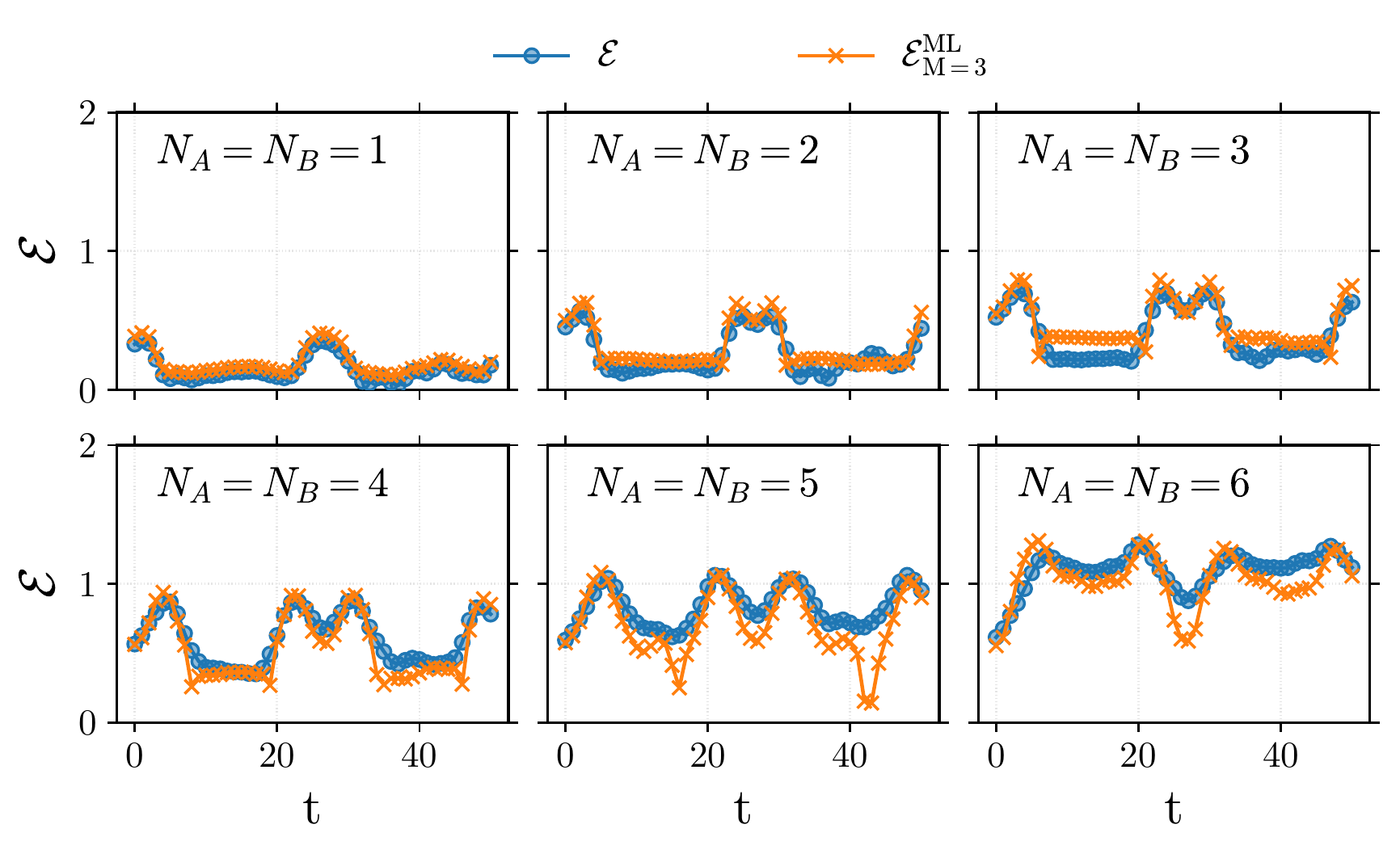}
    \caption{
    Estimated entanglement when quenching across the Ising phase transition at $B_X=0.5$. The initial state is the groundstate at $B_X = 0.5 + \Delta$, dynamics are generated by quenching with the Hamiltonian at $0.5 - \Delta$, taking $\Delta=0.1$.
    The total size is $L=20$ and the tri-partition is chosen so that subsystems $A$ and $B$ are adjacent and of equal size.
    The blue and orange lines show the true logarithmic negativity and the neural network estimated quantity with 3 copies respectively.
    }
    \label{fig:phase-quench}
\end{figure}

In the main text we quench from the Neel-state (which can be thought of as the ground-states of the XXZ-model with infinite anisotropy) to the isotropic Heisenberg point, which is the location of a Kosterlitz-Thouless phase transition.
This type of quench quickly generates volume-law entanglement, and this is accurately captured by our neural network approach, as show in Fig.~3 of the main text.
One might also consider a different quench, across a different type of quantum phase transition.
Here, we take the paradigmatic example of the transverse field Ising model:
\begin{equation}
    H_{\text{Ising}}(B_X) =
    \sum_{i=1}^{L-1}
    \sigma_{i}^Z \sigma_{i + 1}^Z +
    B_X \sum_{i=1}^{L} \sigma^X_{i},
\end{equation}
where $B_X$ is the magnetic field, which induces a second order phase at the critical point $B_X=0.5$.

In Fig.~\ref{fig:phase-quench} we show the logarithmic negativity as a function of time, exact and predicted from moments using the neural network, for a variety of sizes of two adjacent blocks of spins during a quench across this phase transition.
Specifically, we take the initial state as the ground-state of $H_{\text{Ising}}(1 + \delta)$, and evolve it with $H_{\text{Ising}}(1 - \delta)$.
As before, the spin chain is divided into three blocks, of size $N_A$, $N_B$ and $N_C$.
System $C$ is traced out, then the entanglement is found between $A$ and $B$. This state is generally mixed as long as $N_C > 0$.
The neural network is also the same as before, and we only show results for 3 moments (copies) since this is already sufficient for good accuracy.
This further confirms that the training set chosen is particularly suitable for highly entangled states such as those generated in quenches.

\end{document}